\documentclass[preprint, 11pt,aps]{revtex4}
\usepackage{latexsym}
\usepackage{epsfig}
\usepackage{amsmath}
\usepackage{amssymb}
\usepackage{graphicx}
\usepackage{axodraw}
\usepackage{bbm}
\usepackage{bm}

\usepackage{color}
\definecolor{rosso}{cmyk}{0,1,1,0.4}
\definecolor{rossos}{cmyk}{0,1,1,0.55}
\definecolor{rossoc}{cmyk}{0,1,1,0.2}
\definecolor{blu}{cmyk}{1,1,0,0.3}
\definecolor{blus}{cmyk}{1,1,0,0.6}
\definecolor{bluc}{cmyk}{1,1,0,0.1}
\definecolor{verde}{cmyk}{0.92,0,0.59,0.25}
\definecolor{verdec}{cmyk}{0.92,0,0.59,0.15}
\definecolor{verdes}{cmyk}{0.92,0,0.59,0.4}

\definecolor{rossoCP3}{cmyk}{0,.88,.77,.40}

\makeatletter

\newcommand{\beq}{\begin{equation}}
\newcommand{\eeq}{\end{equation}}
\newcommand{\bea}{\begin{eqnarray}}
\newcommand{\eea}{\end{eqnarray}}
\newcommand{\ba}{\begin{array}}
\newcommand{\ea}{\end{array}}
\newcommand{\bi}{\begin{itemize}}
\newcommand{\ei}{\end{itemize}}
\newcommand{\bn}{\begin{enumerate}}
\newcommand{\en}{\end{enumerate}}
\newcommand{\bc}{\begin{center}}
\newcommand{\ec}{\end{center}}

\newcommand{\no}{\nonumber}
\newcommand{\nn}{\nonumber\\}

\newcommand{\gsim}{\lower.7ex\hbox{$\;\stackrel{\textstyle>}{\sim}\;$}}
\newcommand{\lsim}{\lower.7ex\hbox{$\;\stackrel{\textstyle<}{\sim}\;$}}

\begin{document}
 
\author{Isabella {\sc Masina}$^{\color{rossoCP3}{\clubsuit} \color{rossoCP3}{\heartsuit}}$}
\email{masina@fe.infn.it}
\author{Francesco {\sc Sannino}$^{\color{rossoCP3}{\heartsuit}}$}
\email{sannino@cp3-origins.net}

\affiliation{\vskip 1cm
$^{\color{rossoCP3}{\clubsuit}}$ 
{Dip.~di Fisica dell'Universit\`a di Ferrara and INFN Sez.~di Ferrara, Via Saragat 1, I-44100 Ferrara, Italy}\\
\mbox {$^{\color{rossoCP3}{\heartsuit}}${ {\large C}entre for {\large P}article {\large  P}hysics  {\large  P}henomenology}, {\color{rossoCP3}\large CP}$^{ \bf \color{rossoCP3} 3}${ $\color{rossoCP3}-$ \color{rossoCP3} Origins}}\\  \& \\ 
 {  {\large D}anish  {\large I}nstitute {\large  f}or  {\large  A}dvanced   {\large  S}tudy}, {\color{rossoCP3}\large DIAS},
 \\
{\mbox{University of Southern Denmark, Campusvej 55, DK-5230 Odense M, Denmark.}
}
\vskip 1cm}
%
%
\title{\Large \color{rossoCP3} \bf Cosmic Ray Electron and Positron Excesses \\from a Fourth Generation Heavy Majorana Neutrino}

\baselineskip=15pt
\setcounter{page}{1}

\vskip 2.5cm

\begin{abstract}
Unexpected features in the energy spectra of cosmic rays electrons and positrons 
have been recently observed by PAMELA and Fermi-LAT satellite experiments, 
opening to the exciting possibility of an indirect manifestation of new physics. 
A TeV-scale fourth lepton family is a natural extension of the Standard Model leptonic sector 
(also linked to the hierarchy problem in Minimal Walking Technicolor models). 
The heavy Majorana neutrino of this setup mixes with Standard Model charged leptons 
through a weak charged current interaction.
Here, we first study analytically the energy spectrum of the electrons and positrons originated
in the heavy Majorana neutrino decay modes, also including polarization effects. 
We then compare the prediction of this model with the experimental data, exploiting both the 
standard direct method and our recently proposed Sum Rules method.  
We find that the decay modes involving the tau and/or the muon charged leptons as primary decay products 
fit well the PAMELA and Fermi-LAT lepton excesses while there is tension with respect to the antiproton to proton fraction constrained by PAMELA. 
\\[.1cm]
{\footnotesize  \it Preprint: CP$^3$-Origins-2011-15\\
{\footnotesize  \it Preprint: DIAS-2011-01}}
\end{abstract}

\maketitle
\tableofcontents

\newpage
\section{Introduction}

A hardening in the electron and positron fluxes has been observed in Cosmic Rays (CRs) at energies between 
$10$ GeV and $1$ TeV. More precisely, PAMELA \cite{Adriani:2008zr} measured a rise in the positron fraction above
$10$ GeV. Fermi-LAT \cite{Abdo:2009zk} measured the sum of the positron and electron fluxes, detecting a feature 
with spectral index of about -3 above $100$ GeV, followed by a decrease at $500$ GeV.
The latter results strengthened previous hints by ATIC \cite{:2008zzr} and PPB-BETS \cite{Torii:2008xu}
balloon experiments and by the ground-based telescope HESS \cite{Aharonian:2008aa}. 

When compared with the predictions of current astrophysical models, the results of PAMELA and Fermi-LAT
independently suggest the presence of an excess of electrons and/or positrons. 
The origin of these excesses is unknown and many explanations have been put forward: 
from misunderstood astrophysical effects to new physics indirect effects.
Notice that the lack of any excess in antiproton CRs found by PAMELA \cite{Adriani:2010rc} 
is a crucial element in understanding the origin the electron and positron excesses.
Interpreting these excesses as signatures of a dark matter component, generically an 
annihilating or decaying WIMP,
is for sure exciting; see {\it e.g.} \cite{Fan:2010yq} for a review. 

While the interpretation in terms of dark matter annihilations often leads to an unobserved excess 
of gamma and radio photons, the interpretation in terms of dark matter decays is compatible with photon 
observations \cite{Nardi:2008ix,Ibarra:2009nw}, even though some channels now start to show some tension 
\cite{Cirelli:2009dv,Meade:2009iu,Papucci:2009gd,Hutsi:2010ai}.
A discrimination strategy was proposed in \cite{Boehm:2010qt,PalomaresRuiz:2010uu}.

In this work, we want to investigate a simple extension of the Standard Model (SM): 
a heavy Majorana neutrino $N$ which couples to SM particles only via charged current interaction:
\begin{equation}
{\cal L}^{CC}_{\ell N}=-i\frac{g}{\sqrt{2}} ~C_{\ell} ~W_\mu^- \bar \ell \gamma^\mu P_L N + h.c.
\label{vertex}
\end{equation}   
where $C_{\ell}$, with $\ell=e,\mu,\tau$, is a numerical factor parametrizing the strength of the mixing 
between the heavy neutrino $N$ and the SM charged lepton $\ell$ and $P_L$ is the L-chirality projector. 
The interaction above naturally arises extending the SM leptonic sector by introducing a fourth lepton family,
as happens {\it e.g.} in Minimal Walking Technicolor models \cite{Sannino:2004qp, Dietrich:2005jn,Foadi:2007ue,Frandsen:2009fs,Antipin:2009ks,Andersen:2011yj}.

We are interested in a nearly stable and TeV-scale heavy neutrino, which could constitute a fraction
of the dark matter density, $x_N=\rho_N/\rho_{DM}$. Via its semileptonic decays $N\rightarrow \ell^\pm W^\mp$
-- whose branching ratio (BR) into the specific flavor $\ell$ is proportional to $C_\ell^2$ -- 
the heavy Majorana neutrino could also be a source of electron and positron CRs. 
Our goal is to study whether the electron and positron excesses observed by PAMELA and Fermi-LAT
can be attributed to such decaying heavy Majorana neutrino. Notice that in our model the fluxes of electrons 
and positrons coming from the $N$ decay chains are equal.

We focus on a very specific model: we are thus able to discard the model if it turns out not to fit the data, 
or to strongly constrain its parameter space.
Instead of using numerical codes \cite{Ibarra:2008qg,Ibarra:2008jk,Ibarra:2009dr,Nardi:2008ix,Meade:2009iu,Cirelli:2010xx}, 
for each lepton flavor $\ell$ here we calculate analytically the energy spectra of the electrons and positrons 
produced in $N\rightarrow \ell W$ and the subsequent $\ell$ and $W$ decay chains, including polarization effects. 
This allows to have a deeper understanding of the physical results and, as we are going to discuss,
to find features that are not properly accounted for in some numerical codes. 
We then consider the propagation of electrons and positrons according to the most popular models and 
compare our predictions with the experimental results of PAMELA and Fermi-LAT,
exploiting both the more straightforward Sum Rules (SR) method proposed in ref.\cite{Frandsen:2010mr} and the longer direct 
comparison method.   

Considering in turn each channel $N\rightarrow \ell W$ (with $\ell=e,\mu,\tau$) and its subsequent decays,
we find that: 
the decays into $\tau W$ fit the experimental data extremely well for $M_N\approx 3$ TeV and 
$x_N^{1/2}C_\tau \approx 10^{-27}$;
a good fit is obtained for the decays into $\mu W$ for the same model parameters; 
while it is impossible to reproduce the data if $N$ decays mostly into $e W$.

As for the CR antiproton contribution from the heavy Majorana neutrino decays, in ref.\cite{Ibarra:2009dr}
it was estimated to be compatible with the PAMELA experimental data. In ref. \cite{Nardi:2008ix} 
there was some tension with the data, but a fit to the ATIC data were used there. We have redone the analysis for the antiproton flux as well as the antiproton to proton fraction for the processes relevant here and compared to CAPRICE \cite{Boezio:2001ac} and PAMELA \cite{Adriani:2010rc} data. We find that it is possible to accommodate the CAPRICE data while we observe a tension with the PAMELA data.

We conclude that a $3$-TeV fourth family Majorana neutrino with dominant BR into $\tau W$ and/or $\mu W$
with coupling $x_N^{1/2}C_{\tau/\mu} \approx 10^{-27}$ represents a plausible and even quite conservative
explanation of the electron and positron CRs excesses.

Our findings confirm previous numerical studies for the $e$ and $\mu$ channel,
but disagree for the $\tau$ channel, which was considered to be disfavored \cite{Ibarra:2009dr}
or to provide a worse fit to the data as compared to the $\mu$ channel \cite{Meade:2009iu}. 
These analysis were based on the Monte Carlo simulation program PYTHIA which (as well as HERWIG)
treats leptons and vector bosons as unpolarized. To understand the origin of this discrepancy, we carried 
out our analysis also neglecting polarization effects. We find that, although going in the right direction,
polarization effects cannot alone explain the discrepancy between our analytically-based results and the 
numerically-based results of previous analyses.  
A comparison with ref. \cite{Nardi:2008ix} is not strictly possible, since polarization effects were neglected 
and especially the ATIC data \cite{:2008zzr} were used for the fits, not the subsequent 
ones by Fermi-LAT \cite{Abdo:2009zk}. On the other hand,
in the more recent analysis of ref.\cite{Cirelli:2010xx} including polarization and even electroweak 
emission \cite{Ciafaloni:2010ti}, the semileptonic decay channels were not considered.

The paper is organized as follows.
In section II the model setup is introduced. In section III the analytical calculation of the 
energy spectrum of electrons and positrons from $N$ decays is carried out, studying separately each
of the three semileptonic channels. Section IV is devoted to the comparison of the model prediction 
with the experimental data, bot with the SR and direct comparison methods. Section V presents
the estimate for the antiproton flux and antiproton-proton ratio. 
We draw our conclusions in section VI.


\section{The fourth lepton family setup}

In the SM, the three lepton families, $\ell=e,\mu,\tau$, belong to the following representations 
of the gauge group $SU(3)_c\times SU(2)_L \times U(1)_Y$:
\beq
L_\ell = ({\nu_\ell}_L ~~ \ell_L)^T \sim (1,2,-1/2) ~~~~~,~~~~~\ell_R \sim (1,1,-1)~~~~~.
\eeq  
where the chirality projectors $P_L=(1-\gamma_5)/2$ and $P_R=(1+\gamma_5)/2$ have been introduced. 
It has been observed experimentally that at least two of the SM neutrinos have a small mass, 
not larger than the eV-scale \cite{pdg2010}. In the following, we account for light neutrino masses and mixings 
via an effective Majorana mass term, i.e.  by adding to the SM Lagrangian a dimension-5 non-renormalizable 
operator (which could arise for instance through a seesaw mechanism). 

Our aim here is to investigate the CR positron and electron fluxes originated via the decay 
of a possible TeV-scale fourth lepton family, for which we introduce the $\zeta$-flavor. 
This additional family is composed by a lepton doublet, a charged lepton singlet and a gauge singlet:
\beq
L_\zeta= ({\nu_\zeta}_L ~~ \zeta_L)^T \sim (1,2,-1/2)~~~~,~~~~~{\zeta}_R\sim (1,1,-1)~~~,
~~~~~{\nu_\zeta}_R\sim (1,1,0)~~. \label{newLeptons}
\eeq 
We keep our scenario as general as possible, but we stress that it would arise naturally in Minimal 
Walking Technicolor models \cite{Sannino:2004qp}, with interesting collider signatures \cite{Frandsen:2009fs} 
- see also \cite{Antipin:2009ks} and, for a complete review, ref.\cite{Andersen:2011yj}.

The $\zeta$-charged lepton, $\zeta = {\zeta}_L+{\zeta}_R$, has a Dirac mass term like the other 
three charged leptons of the SM, but large enough to avoid conflict with the experimental limits.  
We work in the basis in which the $4\times4$ charged lepton mass matrix is diagonal. The mass eigenstates of the
two electrically neutral Weyl fermions instead correspond to two Majorana neutrinos, $N_{1,2}$, for which
we are free to choose the ordering $M_{N_1}\le M_{N_2}$.
In such a framework, assuming the $\zeta$ lepton to be heavy enough with respect to $N_1$,
it turns out that the latter is stable and can constitute - at least part of - the dark matter in our galaxy. 

We consider the possibility that the new heavy leptons mix with the SM leptons.
For clarity of presentation, we assume that the heavy neutrinos mix only with one SM neutrino of flavor 
$\ell$ ($\ell=e,\mu,\tau$). This is not a restrictive assumption: as demonstrated in
appendix~\ref{sec:full mixing}, the expressions that we are going to derive apply also to the case
in which the heavy neutrinos mix simultaneously with all three light lepton families. 
The entries of the mass matrix are:
\begin{equation}
-{\cal L}_{mass}= \frac{1}{2} ( \begin{array}{ccc} 
 \overline{(\nu_{\ell L})^c}& \overline{(\nu_{\zeta L})^c} & \overline{\nu_{\zeta R}}  \end{array} )
\left( \begin{array}{ccc}  {\cal O}(eV) & {\cal O}(eV) & m_\ell \cr  {\cal O}(eV) &  {\cal O}(eV) & m_D \cr 
m_\ell & m_D &  m_R \end{array} \right)  
\left( \begin{array}{c} \nu_{\ell L} \cr \nu_{\zeta L} \cr (\nu_{\zeta R})^c  \end{array} \right) + h.c.~~.
\label{generalmatrix}
\end{equation} 
The measured values of the light neutrino masses suggest that the entries of the upper 2$\times$2 block have 
to be of ${\cal O}(eV)$. The mass scale of $m_R$ could be as large as the cutoff of the theory, since it is 
not protected by any symmetry (we allow violation of the lepton flavour number $L_\zeta$ by 2 units). 
The elements $m_\ell$ and $m_D $ are expected to be at most of the order of the electroweak scale. 
Given such a hierarchical structure and up to small corrections of ${\cal O}(eV/M_{N_1,N_2}) \lesssim 10^{-11}$,
one obtains the following form for the unitary matrix which diagonalises eq.(\ref{generalmatrix}):
\begin{equation}
\left( \begin{array}{c}
\nu_{\ell L} \cr  \nu_{\zeta L} \cr  (\nu_{\zeta R})^c  \end{array} \right) =  V
\left( \begin{array}{c} N_{\ell L} \cr N_{1 L} \cr N_{2 L}  \end{array} \right) ~~,~~
V=\left( \begin{array}{ccc} \cos \theta_\ell & i \cos\theta \sin\theta_\ell & \sin\theta \sin\theta_\ell 
\cr -\sin\theta_\ell & i \cos\theta \cos\theta_\ell & \sin\theta \cos\theta_\ell \cr 
0 & -i \sin\theta & \cos\theta \end{array} \right).
\label{Eq: mixing matrix}
\end{equation}
$N_{\ell,1,2}$ are the new Majorana mass eigenstates (actually $N_\ell$ is a flavor eigenstate
when including all three SM leptons, as discussed in the appendix) and
\begin{equation}
\tan \theta_\ell = \frac{m_\ell}{m_D} ~~~, ~~~~~\tan 2 \theta= 2 ~\frac{\sqrt{m_D^2+m_\ell^2}}{m_R} ~~~.
\label{eqtheta}
\end{equation}
The light neutrino $N_\ell$ has a mass of ${\cal O}(eV)$. Up to corrections of ${\cal O}(eV)$, 
the heavy neutrinos $N_{1,2}$ have positive masses given by: 
\beq
M_{N_1}=\frac{m_R}{2} \left( \sqrt{1+4 \frac{m_D^2+m_\ell^2}{m_R^2}} -1 \right) ~~,~~~~
M_{N_2}=\frac{m_R}{2} \left( \sqrt{1+4 \frac{m_D^2+m_\ell^2}{m_R^2}} +1 \right) ~.
\label{eq:M1M2}
\eeq
The smaller is $m_R$ (namely the more $\theta$ approaches $\pi/4$), the more the neutrinos $N_1$ and $N_2$ 
become the two Weyl components of a Dirac state. 
Notice that the mixing between the heavy neutrino states $N_{1,2}$ is controlled by the mixing angle 
$\theta_\ell$, which is constrained experimentally to be small (more on this later).

The neutral current interactions of the neutrinos are 
\begin{equation}
{\cal L}^{NC}_N=
\frac{g}{4 \cos \theta_w}Z_\mu( \bar N_\ell \gamma^\mu \gamma_5 N_\ell 
+ \cos^2 \theta~ \bar N_1 \gamma^\mu \gamma_5 N_1 
+ \sin^2 \theta~\bar N_2 \gamma^\mu \gamma_5 N_2 
- i \sin(2\theta)~  \bar N_2 \gamma^\mu  N_1) \ ,
\label{eq:NCfull}
\end{equation}
while their charged current interactions with the light and heavy charged leptons are respectively
\beq
{\cal L}^{CC}_{\zeta N} = -\frac{g }{\sqrt{2}} W_\mu^-  \bar \zeta_L \gamma^\mu 
(-\sin\theta_\ell N_{\ell L} + i \cos\theta \cos\theta_\ell N_{1 L} + \sin\theta \cos\theta_\ell N_{2 L})+ h.c.
\label{eq:CCfullzeta}
\eeq 
and
\beq
{\cal L}^{CC}_{\ell N} =- \frac{g }{\sqrt{2}} W_\mu^- \bar \ell_L \gamma^\mu 
(\cos\theta_\ell N_{\ell L} + i \cos\theta \sin\theta_\ell N_{1 L} + \sin\theta \sin\theta_\ell N_{2 L})+ h.c. ~~.
\label{eq:CCfullell}
\eeq
Notice that the neutrino neutral current remains flavor diagonal at tree-level 
(the neutral current is not flavor diagonal in models with TeV scale right handed neutrinos 
involved in the see-saw mechanism for the light SM neutrino masses, see e.g. \cite{del Aguila:2007em}), 
hence the heavy neutrinos couple to the SM ones only through the charged current interactions at this order.
This is a distinctive feature of our fourth lepton family. 

Indeed, also the Yukawa interactions do not mix the heavy Majorana neutrinos with the SM one:  
\bea
{\cal L}^{H}_{N}= - \frac{g~( \cos\theta_\ell m_D +\sin\theta_\ell m_\ell) }{2 M_W}  
\left(\frac{\sin(2\theta)}{2}  ({\bar{N}_1}N_1 +{\bar{N}_2}N_2) 
  - i  \cos(2\theta)  {\bar{N}_1}\gamma_5 N_2 \right) H~ .
\label{Eq:Yukawa's}
\eea


In this work we are then concerned with a heavy Majorana neutrino $N_1$, 
which only couples to the SM charged leptons, and with a small mixing angle. 
If such mixing is small enough, $N_1$ can be considered as nearly stable on cosmological time scales.  
We assume $M_\zeta>M_1+M_W$, so that the charged lepton $\zeta$ already disappeared 
as a dark matter candidate because of its decays into $W N_1$.

As for the possible hierarchies between the heavy neutrinos, two main situations arise: 
if $M_{N_2}>M_{N_1}+M_Z$, also $N_2$ already disappeared; if $M_{N_2}<M_{N_1}+M_Z$, $N_2$
is nearly stable like $N_1$ (in the limit of degenerate masses they form a Dirac neutrino). 
From now on, we consider the first scenario.

\subsection{Laboratory constraints}

From collider experiments there are both direct and indirect constraints on heavy leptons.
We refer to \cite{Frandsen:2009fs} for an updated analysis (together with some useful references) 
and here we just summarize the results relevant for the present analysis.

As stressed, we are concerned with a heavy neutrino $N_1$ which is nearly stable on cosmological time scales.
For this setup, a lower bound on the heavy neutrino mass $M_{N_1}$ can be extracted 
from the constraints on the effective number of neutrinos involved in $Z$ decay:
$M_{N_1}\ge 44 (45.4)$ GeV at $2\sigma$ for a Majorana (Dirac) nearly stable heavy neutrino. 
Note however that in the present analysis we are interested in much bigger values of $M_{N_1}$, as will
be discussed in the following sections.

As for the heavy charged lepton $\zeta$, the lower limit on its mass is $102.6$ GeV if it is stable.
This limit is only slightly weakened in the case of a decaying heavy charged lepton, as in our setup. 

The mixing angle $\theta_\ell$ of the heavy leptons with the SM charge lepton $\ell$
is strongly constrained by lepton universality tests: $\sin^2\theta_\ell \le 0.012 (0.15)$ 
at $3\sigma$ for $\ell=e,\mu$ ($\ell=\tau$). However, as we are going to discuss in the following,
much smaller values of $\theta_\ell$ are required in order to have $N_1$ as a nearly stable heavy neutrino.
Hence, at first order in $\theta_\ell$ the mixing matrix $V$ of eq.(\ref{Eq: mixing matrix}) becomes:
\beq
V \approx\left( \begin{array}{ccc} 1  & i \cos\theta \sin \theta_\ell & \sin \theta \sin \theta_\ell \cr 
                             - \sin\theta_\ell & i \cos\theta \cos\theta_\ell & \sin\theta \cos\theta_\ell \cr 
                                    0 & -i \sin\theta       & \cos\theta \end{array} \right).
\label{Vapprox}
\eeq
We emphasize that this structure arises also in the case of three simultaneous mixings with the
three SM lepton families, as discussed in the appendix. 


\subsection{Comological constraints}

It is well known that a stable or nearly stable fourth generation Dirac neutrino is excluded
as the main component of dark matter.
Its vector coupling to the $Z$ results in an unsuppressed annihilation cross section into fermion pairs 
at low momenta, as well as in a large elastic scattering cross section with nucleons. 
The latter strongly constrains the relic density of Dirac neutrinos to be several orders of magnitude 
smaller than the observed dark matter density.

Some of these problems can be avoided when the Dirac pair is split using a Majorana
mass term as in our setup. 
The axial coupling of the $Z$ boson with $N_i$ ($i=1,2$) suppresses its annihilation cross section 
into fermion pairs at low momenta, as well as the elastic scattering cross section with nucleons.
For the vector coupling between $N_1,N_2$ and $Z$, such suppression is not at work. 
By requiring that $M_2-M_1>{\cal O}(100)$ GeV  (the typical momentum transferred 
in dark matter-nucleon scatterings), large inelastic scattering rates are avoided \cite{TuckerSmith:2001hy}. 

However, even in this case the thermal relic density of $N_1$, $\rho_{N_1}$, turns out be significantly 
smaller the dark matter one, $\rho_{DM}$: see {\it e.g.} \cite{Keung:2011zc} for an updated analysis, 
where it is shown that the $N_1$ thermal relic density can reach up to $20\%$ of the observed dark matter density, 
but is below $10\%$ for most of the parameter space. 

We then account for the possibility that $\rho_{N_1}$ is smaller than the dark matter one, $\rho_{DM}$.
It is however reasonable to assume that they have the same (spherically symmetric) profile:
\beq
\rho_{N_1}(r)= x_{N_1} ~\rho_{DM}(r)~~.
\eeq 
In the following we will consider $x_{N_1}$ as a free parameter and adopt for definiteness 
the Navarro Frenk White (NFW) \cite{Navarro:1995iw} profile for the dark matter halo of the Milky Way:
\beq
\rho_{DM}(r)= \rho_0 \frac{1}{(r/r_c)^{\gamma} [1+(r/r_c)^\alpha]^{\frac{\beta-\gamma}{\alpha}}} 
\eeq
where $\alpha=1$, $\beta=3$, $\gamma=1$, $r_c=20$ kpc and $\rho_0=.26 {\rm GeV}/{\rm cm}^3$. 
The latter normalization is chosen in order to have $\rho_{DM}(r_{sun})=0.30 {\rm GeV}/{\rm cm}^3$, 
where $r_{sun}= 8.5$ kpc.


\subsection{Contribution to electron and positron CRs from $N_1$ decay}

We now turn to the study of the production and propagation of the electrons and positrons 
in CRs, considering the contribution from the decay of the heavy Majorana neutrino $N_1$.

The number densities of electrons and positrons per unit energy 
satisfy the transport equation 
\cite{Kamionkowski:1990ty,Strong:1998pw, Baltz:1998xv, Hisano:2005ec, Delahaye:2007fr, Fan:2010yq}:
\beq
\frac{\partial}{\partial t} f_{e^\pm}(E_{e^\pm},\vec r) = K(E_{e^\pm}) \nabla^2 f_{e^\pm}(E_{e^\pm},\vec r) 
+ \frac{\partial}{\partial E_{e^\pm}} [b(E_{e^\pm}) f_{e^\pm}(E_{e^\pm},\vec r)] + Q_{e^\pm}(E_{e^\pm},\vec r)~~,
\label{diffusion}
\eeq
where $E_{e^\pm}$ denotes the electron or positron energy. 
The first term describes the propagation through the galactic magnetic field and the diffusion constant 
is given by $K(E_{e^\pm})=K_0 (E_{e^\pm}/{\rm GeV})^\delta$. For definiteness, in what follows we adopt the 
MED-model parameters for the propagation, namely $K_0=0.0112 {\rm kpc}^2/{\rm Myr}$ and $\delta=0.70$ 
\cite{Delahaye:2007fr}.
Positrons lose energy through synchrotron radiation and inverse Compton scattering on the cosmic microwave
background radiation and on the galactic starlight at a rate $b(E_{e^\pm})=E_{e^\pm}^2/({\rm GeV} \tau_E)$
where $\tau_E=10^{16}$ sec. 
Finally, the last term in the equation above represents the source term. 

The assumption that positrons and electrons are today in equilibrium implies that we
can impose the steady state condition $\partial f_{e^\pm}(E_{e^\pm},\vec r)/\partial t=0$. 
The galaxy is described as a cylinder of radius $R$ and half-thickness $L$, for which we assume the values
$R=20$ kpc and $L=4$ kpc respectively. We impose that the number density 
of positrons and electrons vanishes at the surface of the cylinder. 

The source term must include all possible mechanisms through which $e^\pm$ are produced,
from astrophysics to possible new physics. As for astrophysics, most CR electrons are likely to come from
supernovae remnants, while positrons are mainly produced in hadronic processes when CR protons collide with 
intergalactic hydrogen.  
Clearly, the linearity of the diffusion equation guarantees that its general solution can be expressed
as a sum of the solutions obtained by considering separately each possible source mechanisms.

As already discussed, we assume that $M_2, M_\zeta >> M_1 $, so that $N_2$ and $\zeta$ have already 
decayed into $N_1$. 
On the contrary, $N_1$ could be nearly stable since it mixes only with light leptons (see eq.(\ref{eq:CCfullell}):
\begin{equation}
{\cal L}^{CC}_{\ell N_1}=-i\frac{g}{\sqrt{2}} C_{\ell} W_\mu^- \bar \ell_L \gamma^\mu N_{1L} + h.c.~~,
\end{equation} 
where the coefficient $C_{\ell}=\cos\theta \sin\theta_\ell$ has to be small because so is 
the mixing angle $\theta_\ell$.

Clearly, for electrons and positrons the source term of eq.(\ref{diffusion}) associated to the decay 
of the heavy Majorana neutrino $N_1$ is:
\beq
Q_{e^\pm}(E_{e^\pm},\vec r)= \frac{\rho_{N_1}(\vec r)}{M_{N_1} \tau_{N_1}} \frac{dN(E_{e^\pm})}{dE_{e^\pm}}
\eeq
where $dN/dE_{e^\pm}$ represents the number of $e^\pm$ per unit energy 
that are produced in the decay of $N_1$, $\tau_{N_1}$ is the lifetime of $N_1$ and $\rho_{N_1}$ is its
present relic density. 
We can split the number of $e^\pm$ per unit energy into
\beq
\frac{dN(E_{e^\pm})}{dE_{e^\pm}}= \sum _{\ell=e,\mu,\tau} BR(N_1\rightarrow \ell W)~
\frac{dN^{(\ell W)}(E_{e^\pm})}{dE_{e^\pm}}~~,
\label{eq:dnde}
\eeq
where $dN^{(\ell W)}(E_{e^\pm})/dE_{e^\pm}$ refers to the specific channel $N_1\rightarrow \ell W$ (whose
vertex is proportional to $C_\ell$) and the subsequent decays of $\ell$ and $W$, see fig.\ref{feyn}. 
Introducing the partial lifetime $\tau^\ell_{N_1}=\tau_{N_1}/BR(N_1\rightarrow \ell W)$, the source term
can be simply written as a sum of the contributions of the three light lepton flavors:
\beq
Q_{e^\pm}(E_{e^\pm},\vec r)= \sum_{\ell} Q^\ell_{e^\pm}(E_{e^\pm},\vec r)~~,~~ 
Q^\ell_{e^\pm}(E_{e^\pm},\vec r)=\frac{\rho_{N_1}(\vec r)}{M_{N_1} \tau^\ell_{N_1}} \frac{dN^{(\ell W)}(E_{e^\pm})}{dE_{e^\pm}}~~.
\eeq
Since at tree level the Feynman amplitudes for the two charge conjugated decays of $N_1$ are equal, 
the energy spectra of positrons and electrons are also equal: 
$dN^{(\ell W)}(E_{e^+})/dE_{e^+}=dN^{(\ell W)}(E_{e^-})/dE_{e^-}$. 
From now on we will omit the apex $\pm$, unless where necessary.

Clearly, we are interested in $f_{e}(E_e,r,z)$ for $z=0$ and $r=r_{sun}=8.5$ kpc.
The solution of the transport equation at the Solar System, can be formally expressed by
\beq
f_{e}(E_e) =\sum_\ell f^\ell_{e}(E_e)~~,~~
f^\ell_{e}(E_e)=\frac{x_{N_1}}{M_{N_1} \tau^\ell_{N_1}} \int_{E_e}^{M_{N_1}} dE'_e ~G(E_e,E'_e)~
\frac{dN^{(\ell W)}(E'_e)}{dE'_e}
\label{solve}
\eeq
where $x_{N_1} =\rho_{N_1}/\rho_{DM}$ and $G(E_e,E'_e)$ is a Green function 
which contains all the astrophysical dependencies and whose explicit form can be found in \cite{Hisano:2005ec}.
Notice that the Green function is not very sensitive to the choice of the dark matter halo profile, 
since the Earth receives only electrons and positrons created within a few kpc from the Sun, where the different halo 
profiles are very similar.

In the present analysis we adopt the approximation of \cite{Ibarra:2008qg}: 
\beq
G(E,E')=  \frac{10^{16}}{E^2} e^{a+b(E^{\delta-1}-{E'}^{\delta-1})} 
\theta(E'-E) \frac{{\rm sec}}{{\rm cm}^3}
\eeq
where  the coefficients $a$ and $b$ are the
appropriate ones for the MED-model and NFW profile: $a=-1.0203$, $b=-1.4493$.
It was found that this approximation works better than $(15-20)\%$ over the whole range of energies
\cite{Ibarra:2008qg}.


\section{Electrons and positrons from heavy Majorana neutrino decay}

We now turn to calculate the $N_1$ lifetime and the number of electrons and positrons produced
in its decay.

\subsection{Lifetime of $N_1$}

The decay width for $N_1$ is 
\beq
\Gamma_{N_1}=\sum_{\ell=e,\mu,\tau} \Gamma^\ell_{N_1}~~,
~~~~\Gamma^\ell_{N_1}=\Gamma[N_1\to W^+ \ell^-]+ \Gamma[N_1\to W^- \ell^+] ~~,
\eeq
where the two charge conjugated width are equal (at tree level) and, 
in the approximation $m_\ell \ll M_W$, are given by 
\beq
\Gamma[N_1\to W^\pm \ell^\mp] 
= \frac{g^2  C_{\ell}^2}{64\pi}\frac{M_{N_1}^3}{M_W^2}
\left(1-\frac{M_W^2}{M_{N_1}^2}\right)\left(1+\frac{M_W^2}{M_{N_1}^2}-2\frac{M_W^4}{M_{N_1}^4}\right)~~.
\eeq
The branching ratio shave a simple expression in terms of the mixings:
\beq
BR(N_1 \rightarrow \ell W) =\frac{C_\ell^2}{C_e^2+C_\mu^2+C_\tau^2}~~,~~\ell=e,\mu,\tau~~.
\eeq

\begin{figure}[t!]\begin{center}
\includegraphics[width=13cm]{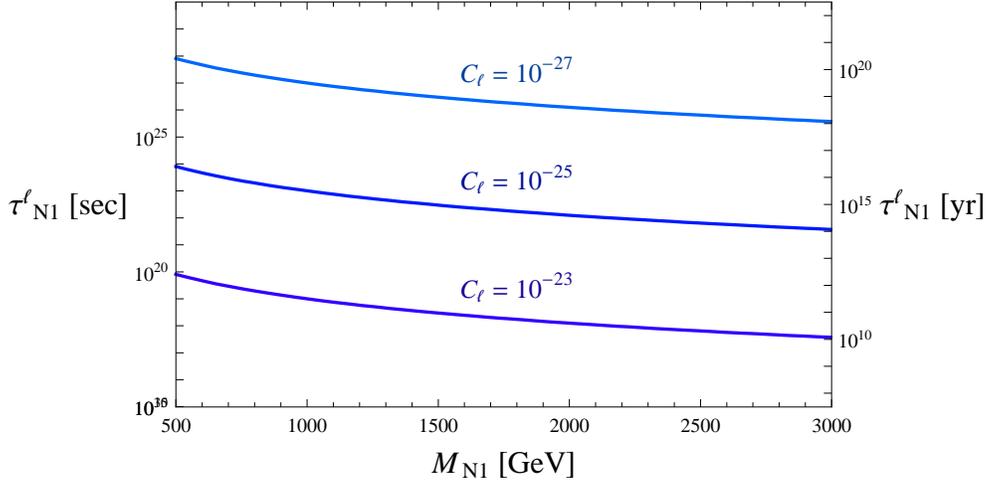}
\end{center}
\vspace*{-0.5cm} 
\caption{Partial lifetime of $N_1$ as a function of its mass $M_{N_1}$ and for selected values of 
the combination of mixing angles $C_{\ell}=\cos\theta \sin\theta_\ell$. }
\label{fig-tau-N1}
\end{figure}

We now consider in turn each decay into a specific lepton flavor $\ell$. 
In fig. \ref{fig-tau-N1} we accordingly display the partial lifetime of $N_1$, $\tau^\ell_{N_1}=1/\Gamma^\ell_{N_1}$,
as a function of its mass and for selected values of $C_{\ell}$, its mixing with the light lepton $\ell$.
We infer that a necessary (not sufficient) condition in oder for the heavy neutrino 
to be stable on cosmological time scales is that all the $C_\ell$'s must be smaller than $10^{-23}$,
so that $\tau_{N_1}^\ell$ is bigger than the age of the universe.

The primary $\ell^{\pm}$ and $W^\mp$, being the decay products of a particle nearly at rest,
are practically monochromatic. In the approximation $m_\ell \ll M_W$, we have
\begin{equation}
p_\ell=E_\ell = \frac{M_{N_1}}{2} \left( 1-\frac{M_W^2}{M_{N_1}^2}\right)=p_W~~~,~~~~ 
E_W = \frac{M_{N_1}}{2} \left( 1+\frac{M_W^2}{M_{N_1}^2}\right)
\label{2E}
\end{equation}
where $p$ and $E$ denote particle's momentum and energy, respectively.

The decaying heavy neutrinos $N_1$ are unpolarized, so the lepton $\ell$ and the $W$ bosons are emitted isotropically.
Focusing on a single decay, it turns out that $\ell^+$ ($\ell^-$) is emitted preferentially 
in the same (opposite) direction as the $N_1$ spin. 
If the decaying $N_1$ has a TeV-scale mass, $\ell^+$ ($\ell^-$) is approximately an helicity
eigenstate with eigenvalue $ +1$ ($-1$). 
The associated $W$ boson can then be considered to be approximately an eigenstate of the 
component of the spin along its direction of motion with $0$ eigenvalue,
that is to be longitudinally polarized.

\subsection{Electron and positron energy spectra}
 
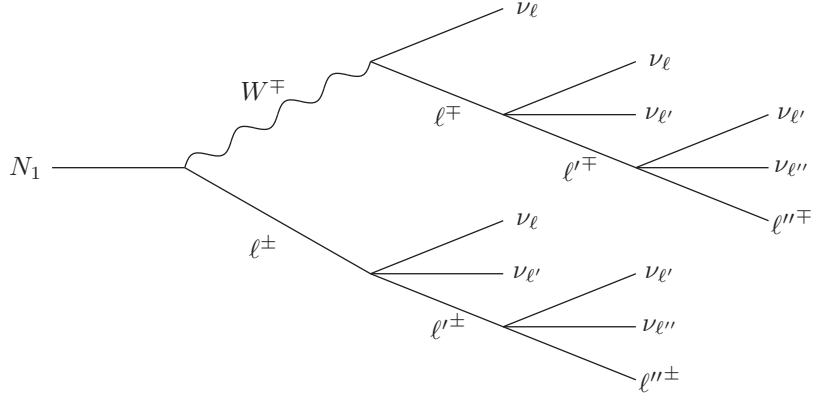
\begin{figure}[t!]
\begin{center} 
\begin{picture}(300,180)(-40,-50)
\Line(-20,60)(30,60) \Text(-30,60)[]{$N_1$}
\Photon(30,60)(100,100){3}{4} \Text(60,90)[]{$W^\mp$} 
           \Line(100,100)(150,120) \Text(160,120)[]{$\nu_{{\ell}}$}
           \Line(100,100)(150,80)  \Text(130,80)[]{${\ell}^\mp$} 
                                                    \Line(150,80)(200,100) \Text(210,100)[]{$\nu_{\ell}$}
                                                    \Line(150,80)(200,80)  \Text(210,80)[]{$\nu_{\ell'}$}
                                                    \Line(150,80)(200,60)  \Text(180,60)[]{${\ell'}^\mp$} 
                                                        \Line(200,60)(250,80) \Text(260,80)[]{$\nu_{\ell'}$}
                                                        \Line(200,60)(250,60) \Text(260,60)[]{$\nu_{\ell''}$}
                                                        \Line(200,60)(250,40) \Text(260,40)[]{${\ell''}^\mp$}           
           
\Line(30,60)(100,20) \Text(60,30)[]{$\ell^\pm$} \Line(100,20)(150,40) \Text(160,40)[]{$\nu_{\ell}$}
                                                \Line(100,20)(150,20) \Text(160,20)[]{$\nu_{\ell'}$}
                                                \Line(100,20)(150,00) \Text(130,0)[]{${\ell'}^\pm$}
                                                      \Line(150,0)(200,20) \Text(210,20)[]{$\nu_{\ell'}$}
                                                      \Line(150,0)(200,0)  \Text(210,0)[]{$\nu_{\ell''}$}
                                                      \Line(150,0)(200,-20)\Text(210,-20)[]{${\ell''}^\pm$}
\label{mm}\end{picture}
\end{center}
\caption{Example of charge conjugated Feynman diagrams for a cascade decay of $N_1$. Clearly, this is just
a suggestive pictorial representation because $m_{\ell}>m_{\ell'}>m_{\ell''}$ is realized only for $\tau,\mu,e$ respectively.}
\label{feyn}
\end{figure}

Clearly, if $\ell^\pm$ is an unstable lepton like $\mu^\pm$ or $\tau^\pm$, it decays 
with a lifetime much shorter than $\tau_{N_1}$, producing $e^\pm$ and neutrinos. 
Also the associated $W^\mp$ boson quickly decays via electroweak interactions.
Since we are interested in a TeV-scale $N_1$, its decay products are highly relativistic particles 
and we can safely work in the narrow-width approximation, {\it i.e.} we treat intermediate particles 
in fig. \ref{feyn} as if they were on-shell.
Accordingly, the number of $e^\pm$ per unit energy obtained from the decay of $N_1$ can be written
as a sum of the contributions from the primary lepton and $W$ boson decay chains:
\beq
\frac{dN^{(\ell W)}(E_{e})}{dE_{e}}=\frac{1}{2}~ 
\left(\frac{dN^{(\ell)}(E_{e})}{dE_{e}}+\frac{dN^{(W)}(E_{e})}{dE_{e}}\right)~~,
\eeq
where the factor $1/2$ has been introduced in order to fix the normalization of the separate 
contributions to one.

\subsubsection{The lepton $\ell$ decay chain}

Let analyse first the electrons and positrons coming from the primary $\ell$ decay chain.

\begin{itemize}

\item[\underline{$\ell = e$}~]
In this case, $e^\pm$ are monochromatic primary decay products of $N_1$, see eq.(\ref{2E}):
\beq
\frac{dN^{(e)}(E_{e})}{dE_{e}} = \delta(E_e - \frac{M_{N_1}}{2} (1-\frac{M^2_W}{M^2_{N_1}}))~~. 
\eeq
This is represented in the top right panel of fig. \ref{fig-dnde} as a vertical line.

\item[\underline{$\ell = \mu$}~]
In this case, $e^\pm$ are produced as secondary decay products from the primary relativistic $\mu^\pm$ 
via its decay into $e^\pm \nu_e  \nu_\mu$. 
They energy spectrum is thus continuous between $0\le E_e \le E_\mu$,
as expected for the 3-body decay of a relativistic particle.
Let take the $z$-axis to be parallel to the muon momentum. As already stressed, the spin of $\mu^+$ ($\mu^-$) 
is parallel (antiparallel) to $\hat z$. 
This means that the $\mu^+$ ($\mu^-$) is polarized in its rest frame, having spin parallel (antiparallel) 
to $\hat z$, so that the angular distribution of the outgoing $e^+$ ($e^-$) is enhanced in the 
same (opposite) direction as the muon spin, namely the $\hat z$ direction:
\beq
\frac{d\tilde \Gamma_{\mu^\pm}}{d\cos\tilde\theta_{e^\pm}} = \frac{\tilde\Gamma_{\mu^\pm}}{2} 
(1 + \frac{1}{3}\cos\tilde\theta_{e^\pm} ) 
\label{thetae}
\eeq
where $\tilde \theta_{e^\pm}$ is the angle between $\hat z$ and the momentum of $e^\pm$ and 
quantities evaluated in the muon rest frame are indicated with a tilde.
A relativistic $e^\pm$ produced in the rest frame of the $\mu^\pm$ with energy $\tilde E_{e^\pm}$, 
is found in the rest frame of $N_1$ to have an energy $E_{e^\pm}= \tilde E_{e^\pm} \gamma (1+\cos\tilde\theta_{e^\pm})$, 
with $\gamma=E_\mu/m_\mu $. 
As a result, the $e^\pm$ emitted with $\tilde\theta_{e^\pm}<\pi/2$ are slightly harder than for an unpolarized $\mu^\pm$
and, assuming $E_\mu \gg m_\mu$, their number per unit energy is given by
\beq
\frac{dN^{(\mu^\pm)}(E_{e^\pm})}{dE_{e^\pm}} = BR(\mu^\pm\rightarrow e^\pm \nu_e \nu_\mu) ~
f_\pm(E_{\mu^\pm},E_{e^\pm},\pm 1)~~,~~0\le E_{e^\pm} \le E_{\mu^\pm}~~,
\label{3Bmu}
\eeq 
where $E_{\mu^\pm}$ is the same as in eq.(\ref{2E}), the BR is close to $100\%$ and the function 
$f_\pm$ is defined as
\beq
f_\pm(E_i, E_f, P_i)= \frac{1}{3 E_i} \left[ 5-9\left(\frac{E_f}{E_i}\right)^2+4\left(\frac{E_f}{E_i}\right)^3 
                         \mp P_i \left(1-9\left(\frac{E_f}{E_i}\right)^2+8\left(\frac{E_f}{E_i}\right)^3 \right)  \right]
\eeq
where $E_i$ and $E_f$ are the energy of the decaying and final lepton respectively and $P_i$
is the polarization of the decaying lepton (see e.g. \cite{Abreu:1995ku}). 
In our case $P_{\mu^\pm}=\pm 1$ and since $f_+(E_i,E_f, 1)= f_-(E_i,E_f, -1)\equiv f(E_i, E_f)$,
it can be directly checked that the energy distributions are equal for electrons and positrons,
as expected. We then omit the apex $\pm$ and write explicitly
\beq
\frac{dN^{(\mu)}(E_{e})}{dE_{e}} = f(E_{\mu},E_{e})~~,~~
f(E_i, E_f)=\frac{4}{3 E_i} \left[ 1- \left(\frac{E_f}{E_i}\right)^3 \right]~~.
\label{dnde-mu}\eeq
The energy distribution $dN^{(\mu)}(E_{e})/dE_{e}$ is represented with a dotted curve in the bottom 
left panel of fig.\ref{fig-dnde}, for $M_{N_1}=1,2,3,5$ TeV from top to bottom.

\item[\underline{$\ell = \tau$}~]
In this case, there are both leptonic and hadronic decay modes,
with BR of about $37\%$ and $63\%$ respectively. 

As for the leptonic decays, secondary $e^\pm$ are produced via $\tau^\pm \rightarrow  e^\pm \nu_e \nu_\tau$
with an energy distribution similar to eq.(\ref{dnde-mu}) but suppressed because of the different BR
\beq
\frac{dN^{(\tau/e)}(E_{e})}{dE_{e}} = BR(\tau\rightarrow e \nu_e \nu_\tau)~
f(E_{\tau},E_{e})~~,~~0\le E_{e} \le E_{\tau}
\label{3Btau}
\eeq 
where $E_{\tau}$ is the same as in eq.(\ref{2E}) and the BR is about $17.8\%$.
The primary $\tau^\pm$ decays also into $\mu^\pm \nu_\mu \nu_\tau$ with a similar BR. 
The latter $\mu^\pm$ has helicity close to $\pm 1$ and fully decays into $e^\pm \nu_e \nu_\mu$, 
giving an additional contribution to the energy distribution 
\begin{equation}
\frac{dN^{(\tau/\mu/e)}(E_e)}{dE_e} = BR(\tau\rightarrow \mu \nu_\mu \nu_\tau)~
\int_{E_e}^{E_\tau} dE_\mu f(E_\tau,E_\mu) f(E_\mu,E_e)
\label{3Btaumu}
\end{equation}   
where $E_\tau$ is as in eq.(\ref{2E}) and $0\le E_e \le E_\tau$.
The sum of the $e^\pm$ energy distributions resulting from $\tau^\pm$ leptonic decays is represented 
with a dash-dotted curve in the bottom right panel of fig.\ref{fig-dnde}, for $M_{N_1}=1,2,3,5$ TeV going 
from top to bottom. Notice that, because of the additional contribution from eq.(\ref{3Btaumu}), 
the dash-dotted curve is softer than the dotted one in the $\mu$ plot at left.

Also the hadronic $\tau^\pm$ decay modes can produce $e^\pm$, which are generically softer than those
produced in its leptonic decay modes. It is not difficult to give an estimate of their 
energy distribution.
The $\tau^\pm$ decay modes into one or two pions are 
$\pi^\pm \nu_\tau$ and $\rho^\pm \nu_\tau \rightarrow \pi^\pm \pi^0 \nu_\tau$, with 
BR$\approx 11\%$ and $25.5\%$ respectively. The $\pi^\pm$ then fully decays into $\mu^\pm \nu_\mu$.
The $\tau^\pm$ decay modes into three or more pions have BR of about $26\%$ and can be neglected
for the sake of the present analysis because the $e^\pm$ produced are very soft.

Let study first the decay $\tau^+ \rightarrow \pi^+ \bar\nu_\tau$ ($\tau^- \rightarrow \pi^- \nu_\tau$).
Since the $\tau^+$ ($\tau^-$) has helicity close to $+1$ ($-1$), in its rest frame it is approximately 
polarized with spin parallel (antiparallel) to $\hat z$, so that the pion momentum is mostly antiparallel to $\hat z$
\beq
\frac{d\tilde \Gamma_\tau}{d\cos\tilde\theta_\pi} = \frac{\tilde\Gamma_\tau}{2} (1 - \cos\tilde\theta_\pi ) ~~,
\eeq
where $\tilde \theta_\pi$ is the angle between $\hat z$ and the pion momentum in the pion rest frame.
The pion mostly goes backwards with respect the the tau momentum and therefore it is softer than for 
an unpolarized tau decay. Indeed in the $N_1$ rest frame
\beq
\frac{dN_\tau(E_\pi)}{dE_\pi}= BR(\tau\rightarrow\pi \nu_\tau)\frac{2}{p_\tau} \left(1-\frac{E_\pi}{E_\tau}\right)~~.
\eeq
The $\pi^+$ ($\pi^-$) decays isotropically in $\mu^+ \nu_\mu$ ($\mu^- \bar\nu_\mu$), so that in the $N_1$ rest frame
the muon has simply a flat energy distribution equal to $1/p_\pi \approx 1/E_\pi$. The $\mu^+$ ($\mu^-$) has 
approximately helicity $+1$ (-1) and fully decays into $e^+ \nu_e\bar\nu_\mu$ ($e^- \bar \nu_e \nu_\mu$). 
The energy distribution of $e^+$ ($e^-$) is finally:
\beq
\frac{dN^{(\tau/\pi/\mu/e)}(E_e)}{dE_e} = \int_{E_e}^{E_\tau} dE_\pi  \frac{dN_\tau(E_\pi)}{dE_\pi}
\int_{E_e}^{E_\pi} dE_\mu \frac{1}{p_\pi} f(E_\mu,E_e)~~.
\eeq

Secondly, we consider the decay $\tau^+ \rightarrow \rho^+ \bar\nu_\tau $ ($\tau^- \rightarrow \rho^- \nu_\tau$).
The $\rho$ having spin $1$, it can take several polarization states and its
angular distribution is less sensitive to the tau polarization 
\beq
\frac{d\tilde \Gamma_\tau}{d\cos\tilde\theta_\rho} = \frac{\tilde\Gamma_\tau}{2} (1 - 0.46 \cos\tilde\theta_\rho ) ~~,
\eeq
but still the $\rho$ preferentially goes backwards with respect to the tau momentum and in the $N_1$ rest frame
\beq
\frac{dN_\tau(E_\rho)}{dE_\rho}= BR(\tau\rightarrow\rho \nu_\tau)\frac{2}{p_\tau} \left(0.73- 0.46\frac{E_\rho}{E_\tau}\right)~~.
\eeq
The $\rho^\pm$ is mildly polarized and it decays nearly isotropically in $\pi^\pm \pi^0$, 
with $\pi^\pm \rightarrow \mu^\pm \nu_\mu$. For the sake of the present analysis 
the electron energy distribution can be approximated by
\beq
\frac{dN^{(\tau/\rho/ \pi/\mu/e)}(E_e)}{dE_e} = \int_{E_e}^{E_\tau} dE_\rho \frac{dN_\tau(E_\rho)}{dE_\rho}
\int_{E_e}^{E_\rho} dE_\pi  \frac{1}{p_\rho}
\int_{E_e}^{E_\pi} dE_\mu \frac{1}{p_\pi} f(E_\mu,E_e)~~.
\eeq

As already remarked, the remaining decays of the $\tau^\pm$ involve three or more pions in the final state: 
the $e^\pm$ eventually produced are even softer and can be neglected in the present analysis.

The sum of the $e^\pm$ energy distributions resulting from $\tau^\pm$ leptonic and hadronic decays is represented 
with a dotted curve in the bottom right panel of fig.\ref{fig-dnde}, for $M_{N_1}=1,2,3,5$ TeV going 
from top to bottom. Clearly, the inclusion of the hadronic tau decay modes has the effect of softening the spectrum.

\end{itemize}

\subsubsection{The $W$ boson decay chain}

Also the primary and approximately linearly polarized $W$ boson further decays.
The number of $e^\pm$ per unit energy produced in its decay chain has to be added 
to the ones studied above.
The $W$ leptonic decay modes into pairs ${\ell} \nu_{\ell}$ have a BR of about $33\%$; 
the remaining $67\%$ is represented by hadronic decay modes. 
Let now identify direction of the $W$ momentum with the $\hat z$ axis.

As for the leptonic decay, in the $W$ rest frame (were the polarization vector is approximately 
$\vec \epsilon = i \hat z$) the secondary $\ell$ (see fig.\ref{feyn}) is emitted with angular distribution given by:
\beq
\frac{d\tilde \Gamma_W}{d\cos\tilde\theta_\ell} = \frac{3\tilde\Gamma_W}{4} (1 -  \cos^2\tilde\theta_\ell ) ~~,
\eeq 
where $\tilde \theta_\ell$ is the angle between the $\hat z$ axis and the $\ell$ momentum.
The lepton is preferentially emitted perpendicularly to $\hat z$ axis in the $W$ rest frame. In the 
$N_1$ rest frame it then displays an energy spectrum peaked at $E_W/2$:
\begin{equation}
\frac{dN_{W}(E_{\ell})}{dE_{\ell}} =BR(W\rightarrow {\ell} \nu_{\ell}) \frac{6}{p_W}~
\frac{E_\ell}{E_W} \left(1-\frac{E_\ell}{E_W}\right) ~~,~~0\le E_\ell \le E_W
\label{decW}
\end{equation} 
where $p_W$ and $E_W$ are given in eq.(\ref{2E}) and we recall that the BR is roughly $11\%$
for all lepton flavors. 
At this point we have to sum over the flavors $\ell=e,\mu,\tau$. 
When $\ell=e$, its energy spectrum can be directly read out from eq.({\ref{decW}}) above:
\beq
\frac{dN^{(W/e)}(E_{e})}{dE_{e}} =BR(W\rightarrow {e} \nu_{e}) \frac{6}{p_W}~
\frac{E_e}{E_W} \left(1-\frac{E_e}{E_W}\right) ~~,~~0\le E_e \le E_W~~.
\label{dnde-We}
\eeq
When $\ell=\mu$, 
the contribution to the $e$ energy spectrum is given by
\begin{equation}
\frac{dN^{(W/\mu/e)}(E_e)}{dE_e} = 
\int_{E_e}^{E_W} dE_{\mu} \frac{dN_{W}(E_{\mu})}{dE_{\mu}} f(E_{\mu},E_e)~~.
\end{equation}
When $\ell=\tau$, there is a similar but suppressed contribution
\begin{equation}
\frac{dN^{(W/\tau/e)}(E_e)}{dE_e} = BR(\tau \rightarrow e \nu_e \nu_\tau)
\int_{E_e}^{E_W} dE_{\tau} \frac{dN_{W}(E_{\tau})}{dE_{\tau}} f(E_{\tau},E_e)~~.
\end{equation}
Since the $\tau$ also decays also into $\mu \nu_\mu \nu_\tau$, there is another small contribution from
\begin{equation}
\frac{dN^{(W/\tau/\mu/e)}(E_e)}{dE_e} = BR(\tau \rightarrow \mu \nu_\mu \nu_\tau)
\int_{E_e}^{E_W} dE_{\tau} \frac{dN_{W}(E_{\tau})}{dE_{\tau}} 
\int_{E_e}^{E_\tau} dE_{\mu} f(E_{\tau},E_\mu) f(E_{\mu},E_e)~~.
\end{equation}
Summarizing, for each primary $W^\pm$, the total number of $e^\pm$ produced 
in its leptonic decay modes is approximately $(11+11+2\times 0.17\times 11) \% \approx  1/4$.

The hadronic decays of the $W$ amounts to about $ 67\%$. They also can produce leptons,
with a soft spectrum that is negligible for the sake of the present analysis. 

In the top left panel of fig.\ref{fig-dnde} we show the number of $e^\pm$ per unit energy
produced by summing all the leptonic decays of the $W^\pm$ (the dotted line is the sole $\ell=e$ contribution), 
which has to be summed to the one coming from the primary $\ell^\mp$ and its eventual decays, as we are now
going to discuss.

\subsubsection{Results}

\begin{figure}[t!]\begin{center}
\includegraphics[width=6.5cm]{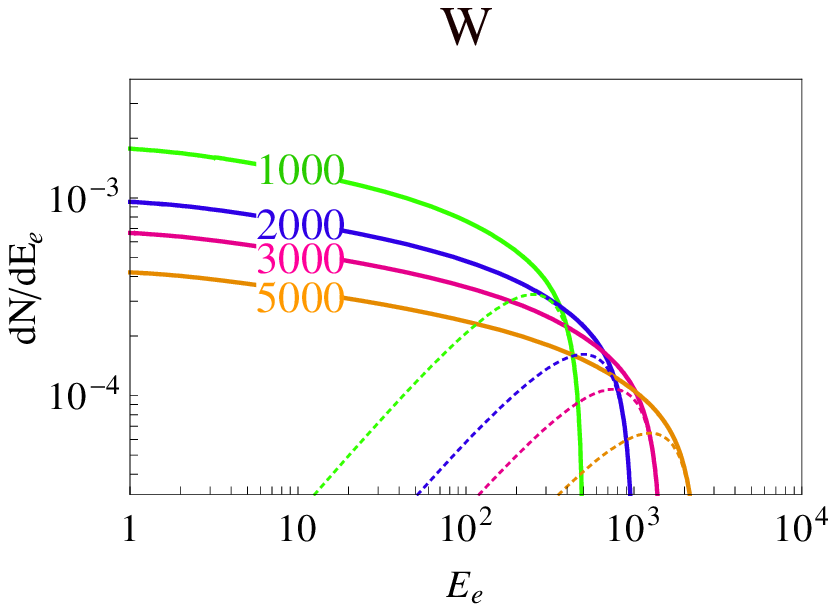}~~~~~~~~\includegraphics[width=6.5cm]{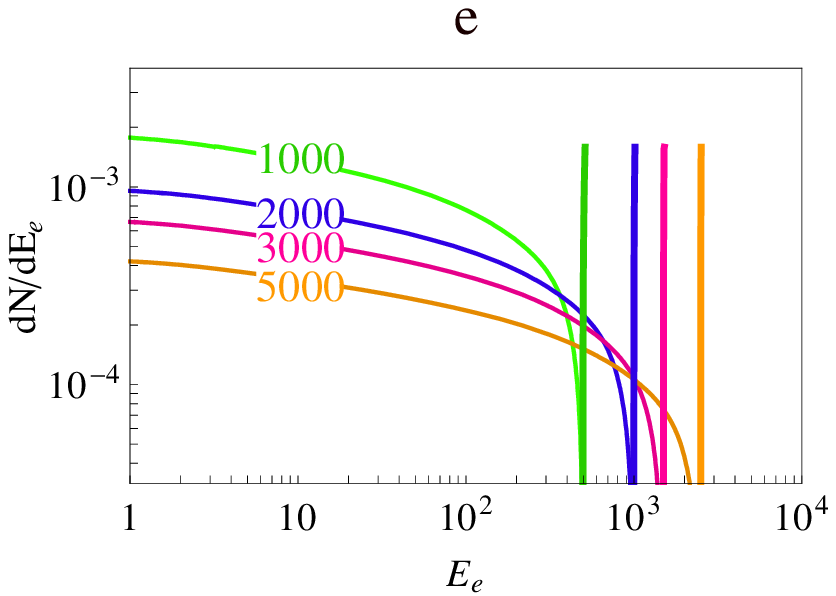}
\vskip.5cm
\includegraphics[width=6.5cm]{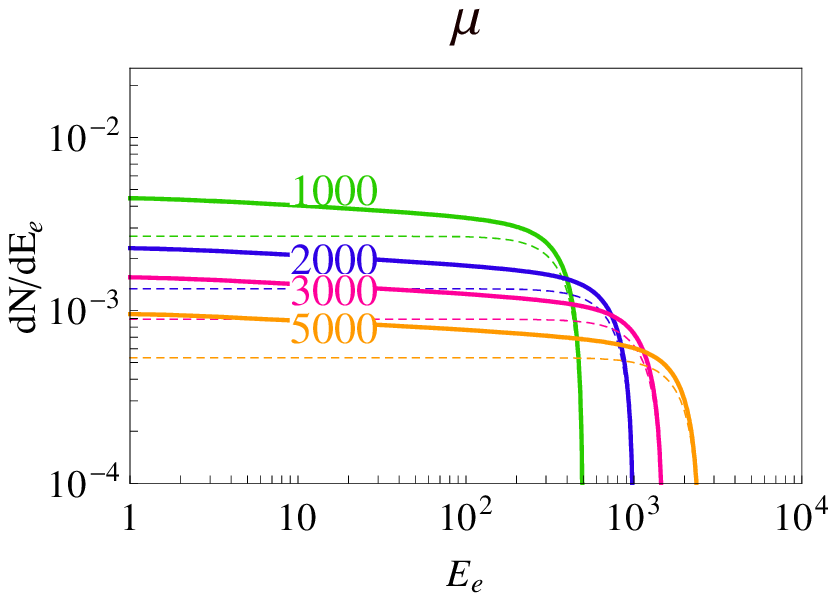}~~~~~~~~\includegraphics[width=6.5cm]{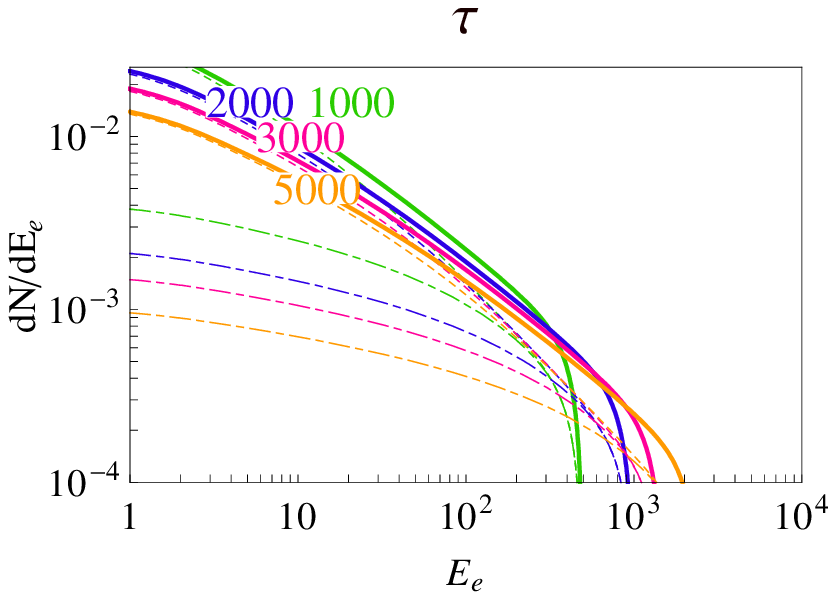}
\end{center}
\vspace*{-0.8cm} 
\caption{Number of $e^\pm$ per unit energy for selected values of $M_{N_1}$ in GeV. 
Top left: the solid lines are the total contribution from the $W$ boson leptonic decay chain; the dotted lines are the
contribution from $W\rightarrow e\nu_e$ only. Top right: solid lines are the sum of the monochromatic $e$ and the
$W$ chain. Bottom left: the solid lines are the sum of the $\mu$ and $W$ chains; the dotted ones are the contribution of 
the $\mu$ chain only. Bottom right: solid lines are the sum of the $\tau$ and $W$ chains; the dash-dotted
lines are the contributions from the leptonic $\tau$ decays; the dotted lines the contribution from
both leptonic and hadronic $\tau$ decays.}
\label{fig-dnde}
\end{figure}

For each lepton flavor, we now sum the energy distributions of $e^\pm$ from the $\ell$ and $W$ boson chains.

For $\ell^\pm=e^\pm$, this sum is easily done since the primary $e^\pm$ are monochromatic, see the top right panel
of fig.\ref{fig-dnde}. 

For $\ell^\pm=\mu^\pm$, the energy spectrum of the secondary $e^\pm$ coming from its decay is soft.
This is shown by the dotted curve in the left bottom panel of fig.\ref{fig-dnde}. 
The solid lines are obtained by adding the contribution of the $e^\pm$ coming from the $W$ chain.
Since the $W$ chain provides approximately $25\%$ of the total number of $e^\pm$
produced in this decay of $N_1$, the contribution is small, especially for the hardest $e^\pm$.  

For $\ell^\pm=\tau^\pm$, the energy spectrum of the $e^\pm$ coming from its leptonic decay is also soft,
but suppressed because of the BR of about $34\%$, as can be seen from the dash-dotted curves in the right 
bottom panel of fig.\ref{fig-dnde}. The dotted line are obtained by adding the contribution from the $\tau$ 
hadronic decay.
Finally, the solid lines are obtained by adding also the contribution from the $W$ chain, which is relevant
at the highest energies and hardens the spectrum close to its end point. 


\subsection{Contributions to $e^\pm$ CR's}

Having obtained explicit expressions for the number of electrons and positrons per unit energy
produced via each channel $N_1 \rightarrow \ell W$ and its subsequent decays, it is straightforward to substitute 
them into eq.(\ref{solve}) and to calculate the flux observed at Earth
\beq
\phi(E_e) = \sum_\ell \phi^\ell(E_e)~~,~~ \phi^\ell(E_e)=\frac{c}{4\pi} f^\ell_e(E_e) ~~,
\eeq
where $c$ is the speed of light. 

\begin{figure}[t!]\begin{center}
\includegraphics[width=6.5cm]{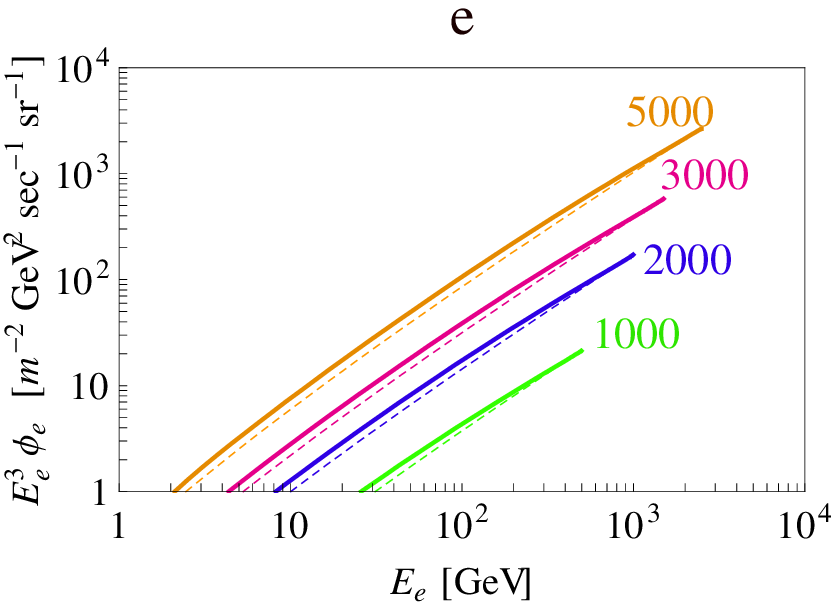}
\vskip.5cm
\includegraphics[width=6.5cm]{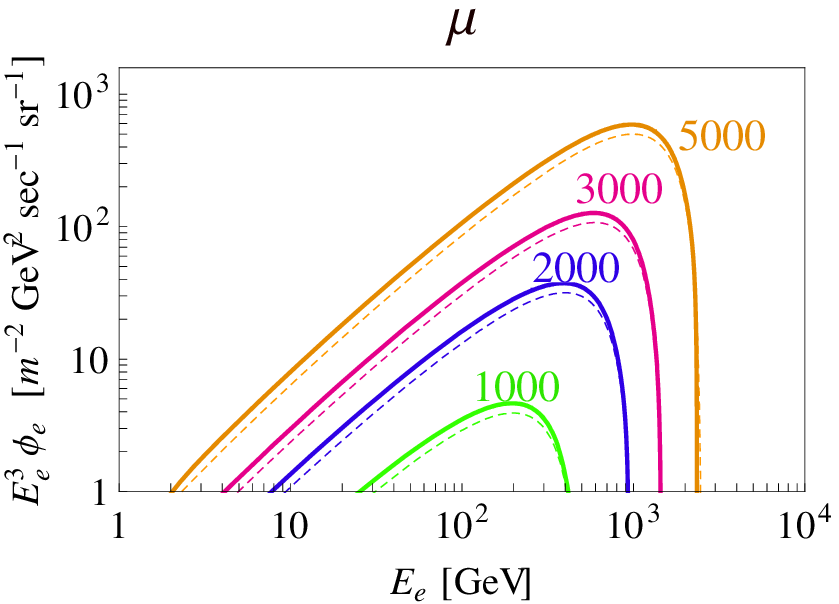}~~~~~~~~
\includegraphics[width=6.5cm]{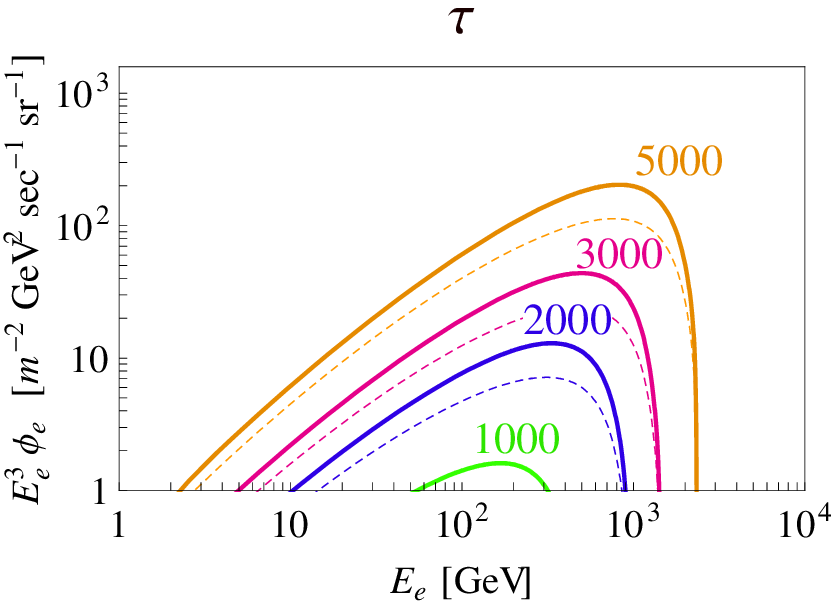}
\end{center}
\vspace*{-0.8cm} 
\caption{Flux of $e^\pm$ multiplied by $E_e^3$ for $N_1\rightarrow \ell W$ with $\ell=e,\mu,\tau$ respectively
and for $M_{N_1}=1000,2000,3000,5000$ GeV. The dashed line is obtained neglecting the contribution from the $W$ chain.
We fixed $C_\ell x_{N_1}^{1/2}=10^{-27}$ everywhere. }
\label{fig-e3phi}
\end{figure}

Clearly, because of the propagation in the galaxy, the energy spectrum of the electrons and positrons gets modified
according to the diffusion equation.
The solid lines in the three panels of fig.\ref{fig-e3phi} display the combination $E_e^3 \phi^\ell(E_e)$ 
for each lepton flavour, 
by assuming $C_\ell x_{N_1}^{1/2}=10^{-27}$ and for four selected values of $M_{N_1}$ in GeV.  
We recall that the fluxes are proportional to $C_\ell^{2}$. 
It is thus not difficult to obtain the total flux $\phi(E_e)$ corresponding to any possible
set of $C_e,C_\mu$ and $C_\tau$.
The dotted line are obtained by neglecting the contribution of electrons and positrons originated from the $W$
boson chain.
The spectrum turns out to be hard for $\ell=e$, while it is softer for $\ell=\mu$, and even more 
for $\ell=\tau$. 

Notice that energy losses due to the diffusion in the galaxy are more relevant at low  
rather than at high $E_e$. This can be realized by comparing, for instance, the plots for the muon case:
the value of the flux increases with $M_{N_1}$ in the whole energy range, while this was not the case
for the number of electrons and positrons per unit energy at the source, see fig.\ref{fig-dnde}. 
The same applies for the tau case, for which the energy spectrum at the source is even softer. 
This justifies the fact that in our calculation we neglected the softer electrons and positrons
coming from the hadronic $\tau$ decays with more than three pions, as well as those coming from the 
hadronic decays of the $W$. Fig.\ref{fig-e3phi} show indeed that the shape of the fluxes after propagation 
is essentially controlled by the behaviour of $dN/dE_e$ close to its end point ({\it i.e.} for the highest possible 
values of $E_e$). For the muon case, this behaviour is a direct reflection of eq.(\ref{dnde-mu}); for the tau,
of the sum of eqs.(\ref{3Btau}),(\ref{3Btaumu}) and (\ref{dnde-We}).   

The same argument can be applied to electroweak radiation effects\cite{Cirelli:2010xx,Ciafaloni:2010ti}. 
Since the mass scale of the Majorana neutrino $N_1$ is much bigger than the electroweak scale,
soft electroweak vector bosons can be radiated in its decay. Such vector boson give
even softer electrons and positrons, they can be neglected in the present analysis because
the cosmic ray excesses observed by PAMELA and Fermi-LAT are comprised in the energy range 
from $20$ GeV to $1$ TeV.

\section{Comparison between the model and the data by PAMELA and Fermi-LAT}

We now come to the comparison between our fourth lepton family model with a decaying TeV-scale Majorana neutrino
and the experimental data on $e^\pm$ CRs recently collected by PAMELA and Fermi-LAT.
We will first describe an extremely fast method, based on the Sum Rules (SR) introduced in ref.\cite{Frandsen:2010mr}.
This method allows for a fast scrutiny of the viability of any model (we considered in particular
the case of models predicting an equal amount of fluxes for electrons and positrons, but the method could of course 
be generalized).
Then, we go through the longer path of the direct comparison, also to show the reliability of the SR method.

\subsection{Sum Rules method}

A fast method to check the viability of our model is to make use of the results of ref.\cite{Frandsen:2010mr},
that we briefly summarize here. 
For definiteness and also for an easy comparison with the literature,
let consider the Moskalenko Strong (MS) \cite{Strong:1998pw, Baltz:1998xv} fluxes $B^\pm (E_e)$, where 
$\pm$ refer to positrons and electrons respectively,
to model the contribution of astrophysical sources (one could of course apply the same reasoning for any other 
astrophysical background model), leaving the normalization of the fluxes $N_B$ as a free parameter:
$\phi_\pm^B(E_e) = N_B B^\pm(E_e)$.  

\begin{figure}[t!]\begin{center}
\includegraphics[width=6.5cm]{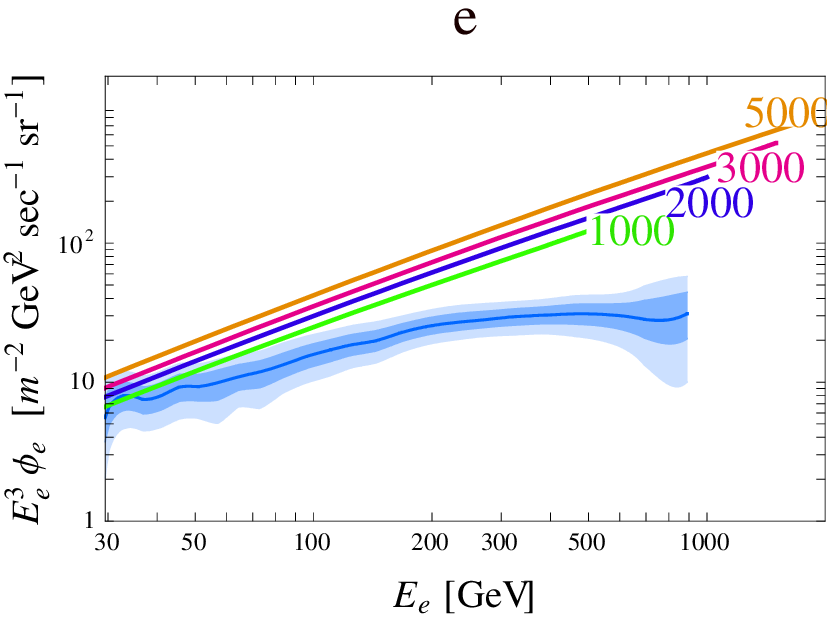}~~~~~~~~
\vskip.5cm
\includegraphics[width=6.5cm]{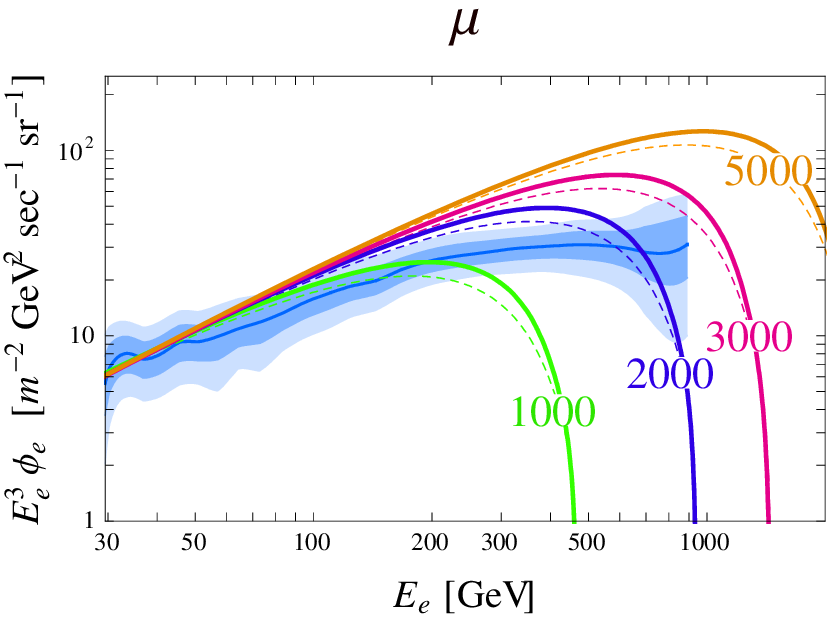}~~~~~~~
\includegraphics[width=6.5cm]{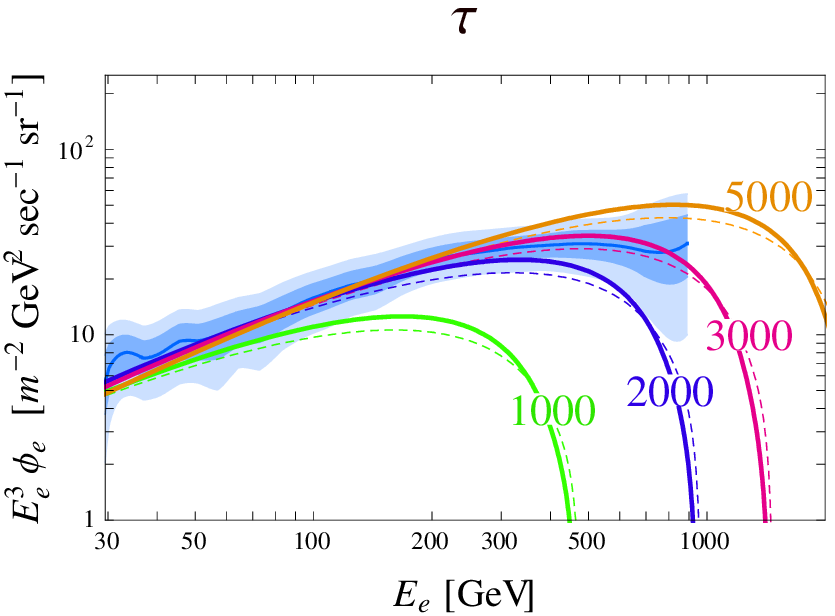}~~~
\end{center}
\vspace*{-0.8cm} 
\caption{Comparison between the prediction of our model for selected values of $M_{N_1}$ in GeV 
and the combined data from PAMELA and Fermi-LAT 
\cite{Frandsen:2010mr}. The inner and outer bands have been obtained by combining respectively
the $1\sigma$ and $2\sigma$ error bands of PAMELA and Fermi-LAT (2010 releases), together with the MS
model for the astrophysical background. For $M_{N_1}=3$ TeV, the curves have been obtained by setting 
$C_e x_{N_1}^{1/2}=10^{-27.02}$,
$C_\mu x_{N_1}^{1/2}=10^{-27.11}$ and 
$C_\tau x_{N_1}^{1/2}=10^{-27.05}$. 
Solid (dashed) lines display the results of including (neglecting) polarization effects.}
\label{fig-sr}
\end{figure}

We assume that an unknown source of $e^\pm$ CR is also present, which gives 
equal fluxes for electrons and positrons, $\phi_+^U(E_{e})=\phi_-^U(E_{e}) \equiv \phi^U(E_e)$.
Hence the total fluxes are
\beq
\phi_{+}(E_e)= \phi^U(E_e)+\phi_+^B(E_e)~~,~~\phi_{-}(E_e)= \phi^U(E_e)+\phi_-^B(E_e)~~.
\eeq 
In ref.\cite{Frandsen:2010mr} we showed how to combine the PAMELA and Fermi-LAT data  
in order to extrapolate the flux of $e^\pm$ from this unknown source.
We re-display this flux $\phi^U(E_e)$ in fig. \ref{fig-sr}, by means of the shaded region. 
The inner and outer bands have been obtained by combining respectively
the $1\sigma$ and $2\sigma$ error bands of PAMELA and Fermi-LAT, while considering all possible
associated values for $N_B$ by means of the SR. 

It is now straightforward to compare the experimentally extrapolated flux $\phi^U(E_e)$ 
with our particular model, belonging indeed to the category
of models which predict equal fluxes for the electrons and positrons. 
All we have to do is to superimpose
the (solid) curves of the fluxes already shown in fig.\ref{fig-e3phi}. 
This is done in fig. \ref{fig-sr}, where the $C_\ell$'s have been fixed so to cross the shaded 
region at the lowest energies.
We recall that the fluxes scale as $C_\ell^{2}$, so that it is not difficult to obtain the value of the
fluxes for any other value of $C_\ell$. The dashed curves have been obtained by neglecting polarization
effects in both $\ell$ and $W$ decay chains. 

We can see that the electron and positron fluxes obtained via the mixing of $N_1$ and the 
$e$-flavor charged lepton have not the right spectral index to fit the combined experimental data.
In the case of mixing with the muon, the electron and positron fluxes follow the shape of the data quite well 
for values of $N_1$ close to $2-3$ TeV. For the mixing with tau, the agreement is even more impressive.
For $M_{N_1}=3$ TeV, the curves have been obtained by setting $C_e x_{N_1}^{1/2}=10^{-27.02}$,
$C_\mu x_{N_1}^{1/2}=10^{-27.11}$ and 
$C_\tau x_{N_1}^{1/2}=10^{-27.05}$.

This means that, if we want to interpret the PAMELA and Fermi-LAT excesses as being due to a decaying heavy 
Majorana neutrino, we need $M_{N_1} \approx 3$ TeV, $C_\tau x_{N_1}^{1/2}\approx 10^{-27}$ and/or
$C_\mu x_{N_1}^{1/2}\approx 10^{-27}$ while $C_e x_{N_1}^{1/2}\lesssim 10^{-28}$. 
Hence the hierarchy $\theta_e < \theta_{\mu,\tau}$
among the mixing angles with the light leptons is needed. 

The universal coupling case $\theta_e =\theta_\mu = \theta_\tau$ is thus excluded, as already pointed out
by the numerical analysis of ref.\cite{Ibarra:2009dr}, where a generic heavy fermion decaying into $\ell W$ 
was considered (the nature of the vertex interaction was not specified).
We postpone to the next section a more complete comparison with previous analysis.


\subsection{Direct method}

We now come to the direct comparison between our fourth lepton family model with a decaying TeV-scale Majorana 
neutrino, considering separately the PAMELA and Fermi-LAT experimental data on $e^\pm$ CRs.

The fluxes of $e^\pm$ arising from $N_1$ decays that we come to calculate have to be summed to those
arising from astrophysical sources. As before,
we consider the popular MS model \cite{Strong:1998pw, Baltz:1998xv} for such astrophysical background
and recall the explicit expressions for $B^\pm(E)$: 
\bea
B^+(E)&=&  \frac{ 4.5 E^{0.7}}{1 + 650E^{2.3} + 1500E^{4.2}} \ ,
 \\
B^-(E)&=& B_1^- + B_2^- \ ,\\ 
B_1^-(E)&=& \frac{ 0.16 E^{-1.1}} {1 + 11 E^{0.9} + 3.2 E^{2.15}}  \ , 
  \\
B_2^-(E)&=& \frac{ 0.70 E^{0.7}}{1 + 110 E^{1.5} + 600 E^{2.9} + 580 E^{4.2}} \ ,
\eea
where $E$ is measured in ${\rm GeV}$ and the $B$s in ${\rm GeV}^{-1} {\rm cm}^{-2}{\rm sec}^{-1}{\rm sr}^{-1}$ units. 
To be specific, in the present analysis we fix the value $N_B = 0.638$, 
obtained by means of the sum rule method proposed in ref.\cite{Frandsen:2010mr}.

\begin{figure}[t!]\begin{center}
\includegraphics[width=6.3cm]{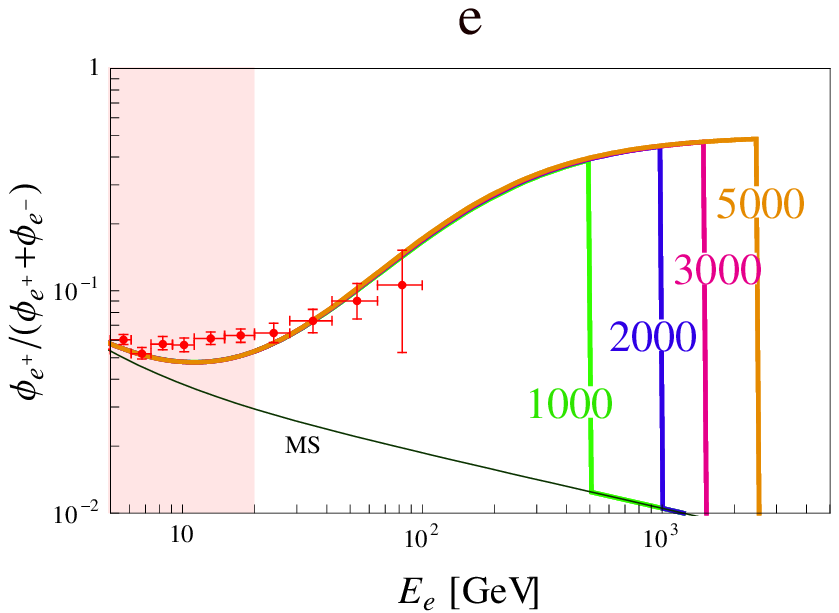}~~~~~~~~\includegraphics[width=6.3cm]{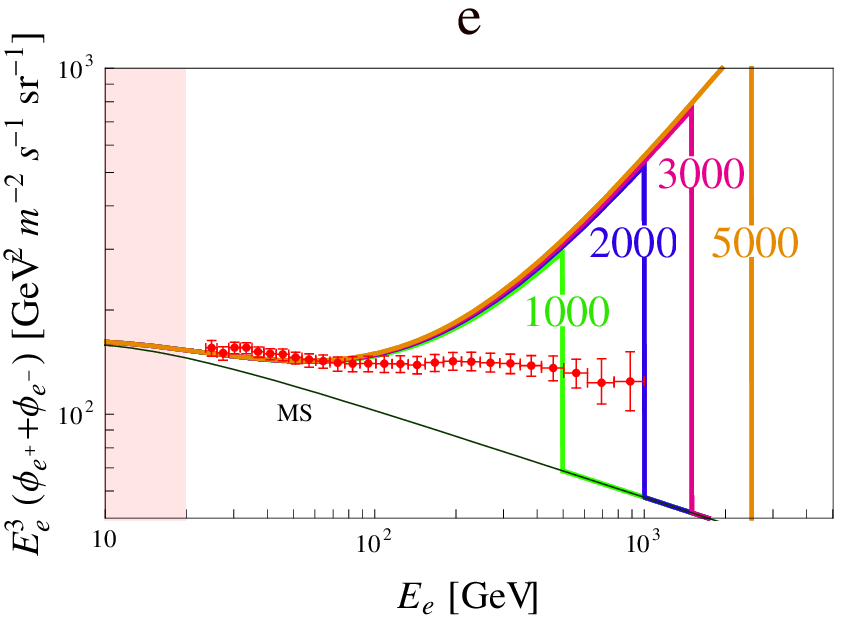}
\vskip.4cm
\includegraphics[width=6.3cm]{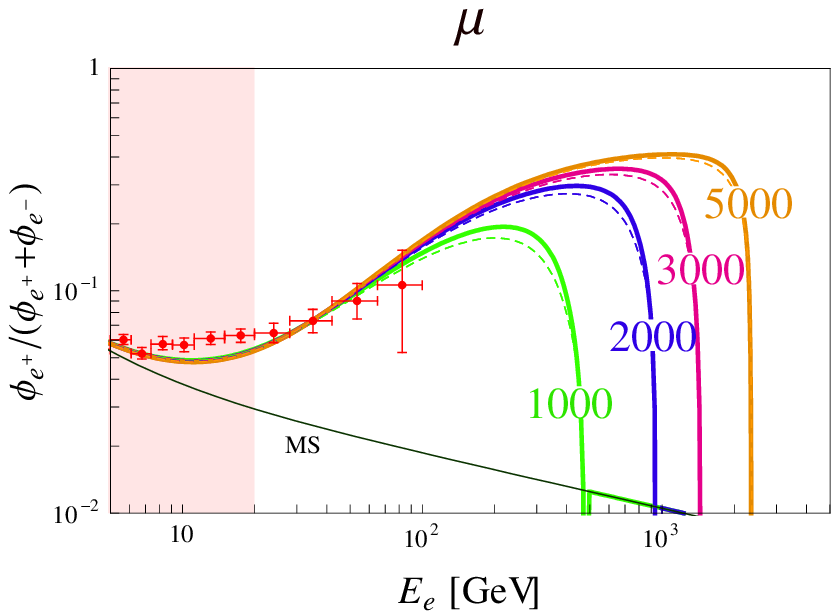}~~~~~~~~\includegraphics[width=6.3cm]{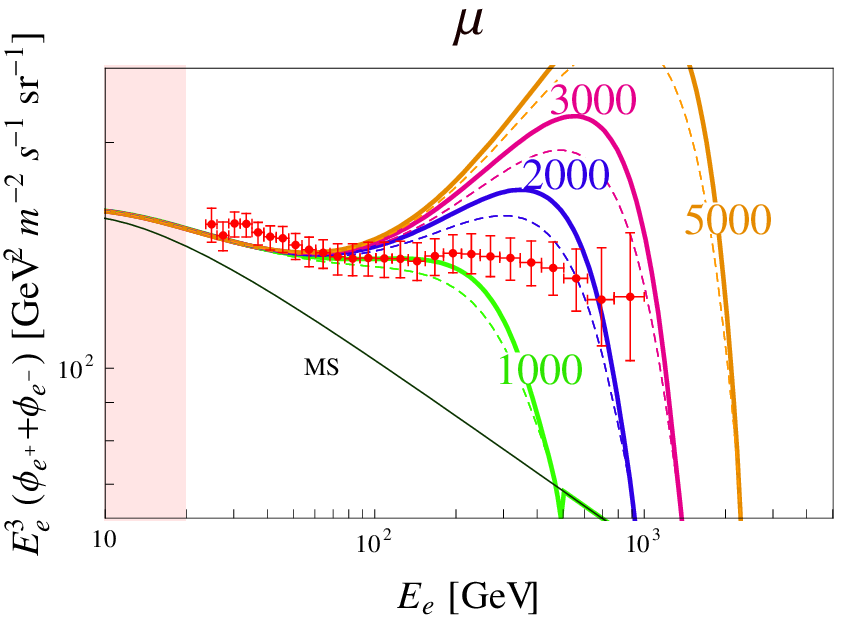}
\vskip.4cm
\includegraphics[width=6.3cm]{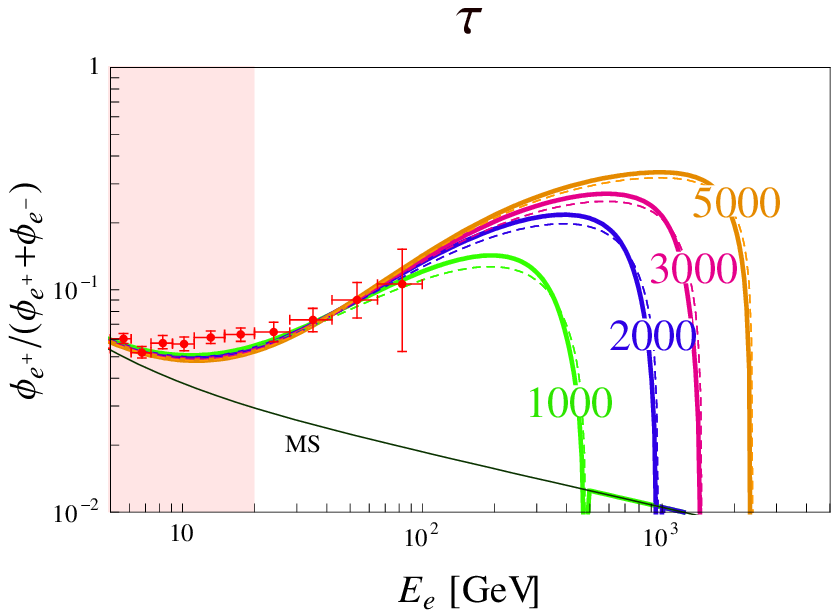}~~~~~~~~\includegraphics[width=6.3cm]{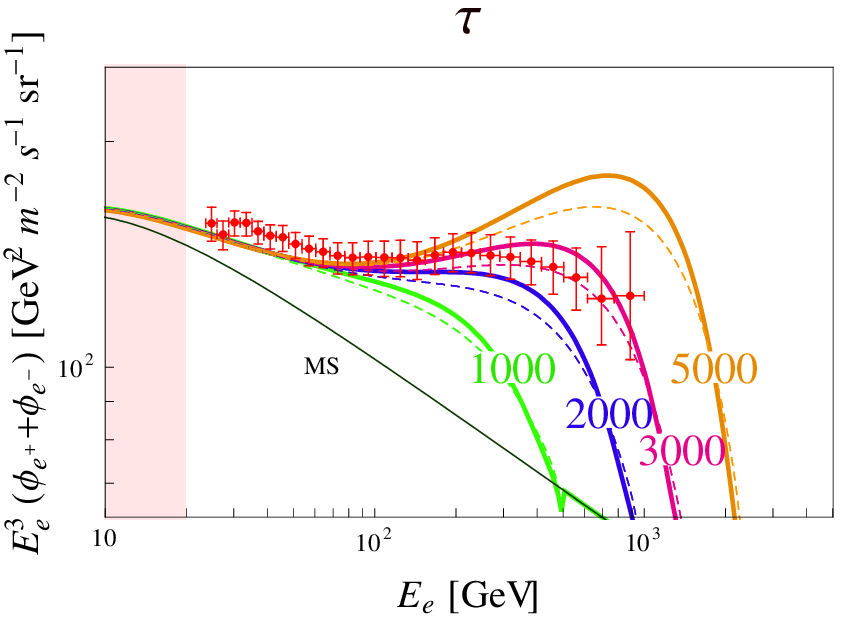}
\end{center}
\vspace*{-0.3cm} 
\caption{Comparison with the 2010 releases of PAMELA (left) and Fermi-LAT 
(right), considering separately the mixing between $N_1$ and each SM lepton $\ell =e,\mu,\tau$. 
The value of the $C_\ell$ fits the central value of the PAMELA $35$ GeV data point.
The solid lines correspond to $M_{N_1} = 1,2,3,5$ TeV from bottom to top respectively.
The shaded region emphasizes the energy range where solar modulation effects cannot be neglected. 
The bottom black line displays the MS estimate for the astrophysical backgrounds. 
For $M_{N_1}=3$ TeV, the values used in the plot for 
$\log_{10}(C_\ell x_{N_1}^{1/2})$ are $-27.105,-27.112,-27.031$ respectively for $e,\mu,\tau$.
Solid (dashed) lines display the results of including (neglecting) polarization effects.}
\label{fig-pf}
\end{figure}

We start by considering the positron fraction and display it in the left panels of fig. \ref{fig-pf} 
for $\ell=e,\mu,\tau$ respectively and for selected values of the $N_1$ mass, $M_{N_1}=1,2,3,5$ TeV. 
Solid (dashed) curves have been obtained by including (neglecting) polarization effects.
The panels also show the PAMELA $1\sigma$ data points (2010 release) and the MS model prediction.
The mixing angle $C_\ell$ has been chosen so that our model curves cross the central value
of the PAMELA positron fraction data point associated to the $35$ GeV energy bin, 
$\phi_{e^+}/(\phi_{e^-}+\phi_{e^+})=0.733$. 
Notice that $C_\ell$ plays the role of shifting up and down the curves with respect to MS.
The shape of the curves depends on the value of $M_{N_1}$, since they drop at about $M_{N_1}/2$. 
The dropping is sharp for $\ell=e$, soft for $\mu$ and even softer for $\tau$. 
This pattern clearly reflects the shape of the fluxes displayed in fig.\ref{fig-e3phi}.
It can be realized that the PAMELA data points are nicely fitted for any value 
of $M_{N_1}$, provided it is bigger than about $200$ GeV \cite{Nardi:2008ix}.

This does not happen for the Fermi-LAT data on the total electron and positron flux, as shown in the right panels
of fig. \ref{fig-pf} for $\ell=e,\mu,\tau$ respectively. The couplings $C_\ell$ have been chosen
to be the same as the corresponding ones in the PAMELA plots at the left.  Again, solid (dashed) curves
have been obtained by including (neglecting) polarization effects.
It is evident that the Fermi-LAT data points cannot be fitted for $\ell=e$, since the curves 
raise too much.  For $\ell=\tau$ the curves have a shape that 
is nicely compatible with the data for values of $M_{N_1}$ close to $2-3$ TeV, while 
for $\ell=\mu$ there is some tension because of an excess around $500-1000$ GeV. 
For $M_{N_1}=3$ TeV, the values of the combination 
$\log_{10}(C_\ell x_{N_1}^{1/2})$ used in the plot are $-27.105,-27.112,-27.031$ respectively for $e,\mu,\tau$.
This means that, if we want to interpret the PAMELA and Fermi-LAT excesses as being due to a decaying heavy 
Majorana neutrino, we need $\theta_e<\theta_{\mu,\tau}$, as concluded in the previous section.

\begin{figure}[t!]\begin{center}
\includegraphics[width=6.5cm]{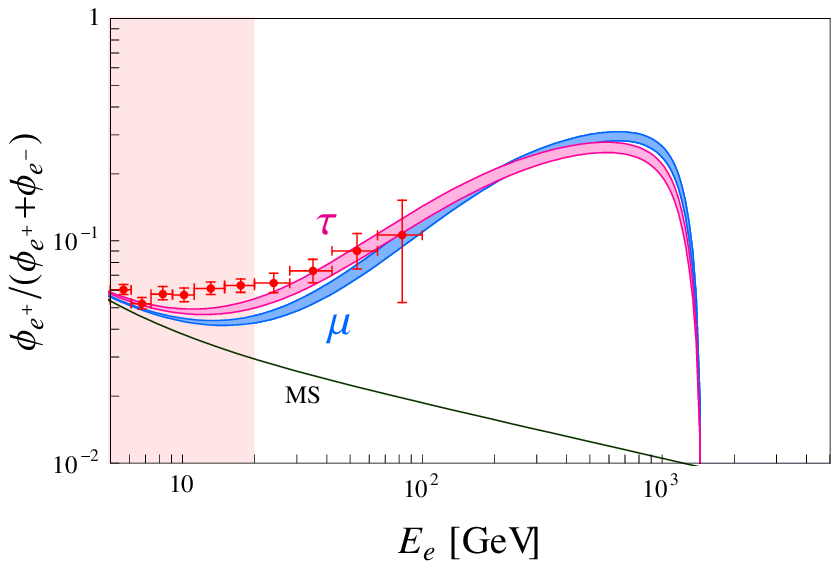}~~~~~~~~\includegraphics[width=6.5cm]{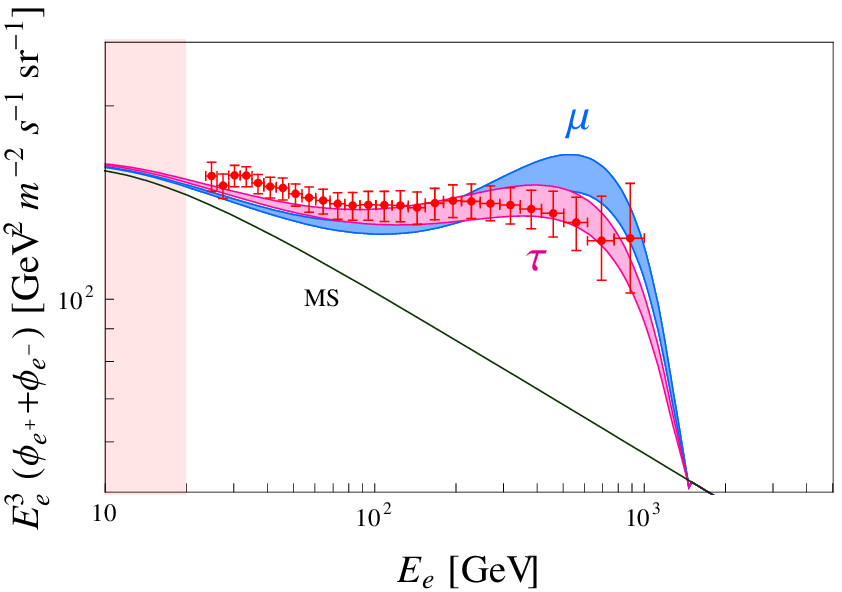}
\put(-345,125){WITH POLARIZATION} \put(-145,125){WITH POLARIZATION}
\vskip.5cm
\includegraphics[width=6.5cm]{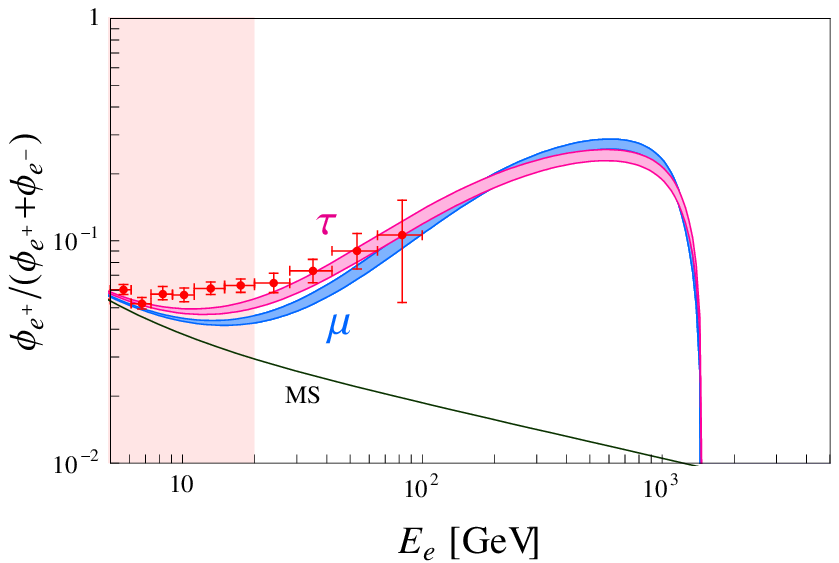}~~~~~~~~\includegraphics[width=6.5cm]{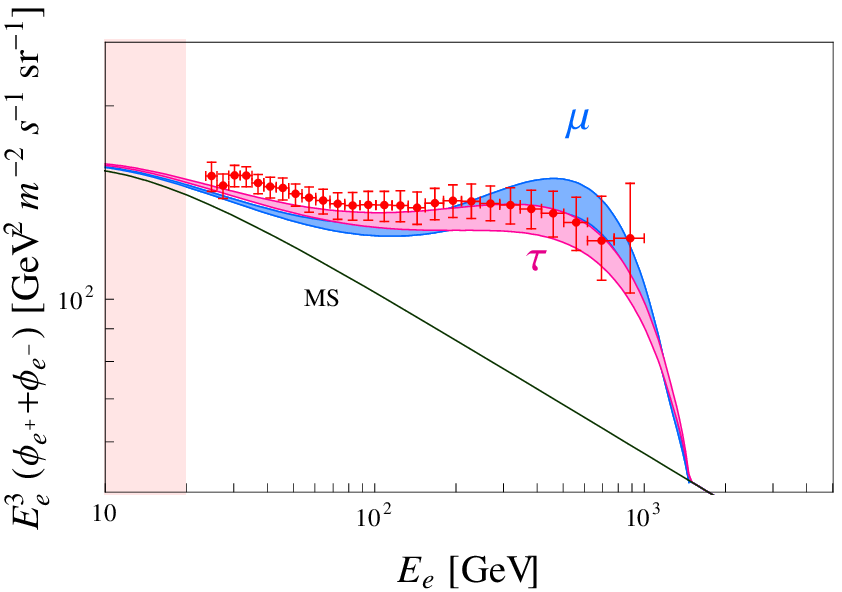}
\put(-350,125){WITHOUT POLARIZATION} \put(-150,125){WITHOUT POLARIZATION}
\end{center}
\vspace*{-0.8cm} 
\caption{Our best fits of PAMELA (left) and Fermi-LAT (right) data.
The mass of $N_1$ has been chosen to be $3$ TeV. In the upper (lower) panels polarization effects have been
included (neglected).
For $N_1\rightarrow \tau W$ we used $x_{N_1}^{1/2} C_\tau=10^{-27.07}-10^{-27.02}$,
while for $N_1\rightarrow \mu W$ we used $x_{N_1}^{1/2} C_\mu=10^{-27.25}-10^{-27.20}$.
In the upper panels polarizations effects are included, in the lower panels they are neglected.}
\label{fig-pf-fine}
\end{figure}

As already stressed, the curve for $\tau$ is slightly softer than for $\mu$ because the its energy spectrum 
is slightly softer, see fig.\ref{fig-dnde} and \ref{fig-e3phi}.
The excess of the $\mu$ curves at $2-3$ TeV could be easily corrected for by slightly reducing 
the coupling $C_\mu$.
This is demonstrated in fig.\ref{fig-pf-fine} where we show the best fit values of $C_\ell$
for $\ell=\mu$ and $\tau$ and for $M_{N_1}=3$ TeV. 
We can see that the $\tau$ fits astonishingly well both PAMELA and Fermi-LAT data
with $x_{N_1}^{1/2} C_\tau=10^{-27.07}-10^{-27.02}$.
The fit for the muon with a slightly suppressed coupling, $x_{N_1}^{1/2} C_\mu=10^{-27.25}-10^{-27.20}$, 
is now good. 
Clearly, we were able to conclude all this also by means of the SR method discussed previously.
However, the direct analysis allows for a better comparison with the results of ref.\cite{Ibarra:2009dr},
as we now turn to discuss. Polarization effects have been included in the upper panels, while they have been
neglected in the lower panels. The inclusion of polarization slightly hardens the total flux.

Ref.\cite{Ibarra:2009dr} concluded that the $\tau W$ channel does not fit the Fermi-LAT data because 
of a too much flat curve; their best fit was obtained in the case of the $\mu W$ channel. 
Indeed, in their plots the fluxes appear softer for both $\mu W$ and $\tau W$ as compared to 
fig.\ref{fig-pf-fine}, even neglecting polarization effects; 
the latter is in particular so soft that the agreement 
with the Fermi-LAT experimental data is lost. In ref. \cite{Meade:2009iu} the $\tau W$ mode was considered to
provide a good fit of the data, but worse than the $\mu W$ mode.
We suspect that the discrepancy with previous numerical analysis 
could either be due to a vertex interaction different from eq.(\ref{vertex}) (for instance
including all chiralities) or to their numerical code. Their analysis is based on PYTHIA, 
that treats leptons and vector bosons as unpolarized. However, as diplayed in fig.\ref{fig-pf-fine},
the inclusion of polarization effects cannot alone explain the discrepancy.

The hardness of the upper right panel spectra for $\mu$ and $\tau$ follows directly from the fact that the CC vertex 
involves L-handed (R-handed) leptons (antileptons). If we had considered a non chiral vertex or if we had neglected 
polarization effects in our calculation, the curves in fig.\ref{fig-pf-fine} would have been softer, as 
demonstrated in the lower right panel.
This can be easily understood for the $\mu$ case, since the contribution from the $W$ decay chain is negligible
and the shape of the flux in fig.\ref{fig-pf-fine} is essentially controlled by the energy distribution of the 
$e^\pm$ produced in the $\mu^\pm$ decay, eq.(\ref{dnde-mu}). For instance, for a relativistic L-handed 
$\mu^{-}$, the highest energy $e^{-}$ produced in its decay are slightly harder than in the unpolarized case
because they are preferentially produced with momentum parallel to the muon one, eq. (\ref{thetae}).
At the contrary, for relativistic R-handed $\mu^-$, the highest energy electrons preferentially escape with
opposite momentum and are thus softer.     
 


\section{Cosmic Ray Antiprotons}

In this section we give an estimate for the cosmic ray antiproton flux resulting from the heavy Majorana 
neutrino decay and compare it with current data.
 
Protons and antiprotons are generated via the hadronic decay of the primary $W$ boson, with BR of about $67\%$.
Their energy spectrum is determined by the fragmentation and hadronization processes. In this case an analytic 
approach is not suitable and we rather adopt the numerical recipies provided in ref. \cite{Cirelli:2010xx}.
In particular, the antiproton (proton) flux obtained from a $1.5$ TeV dark matter annihilating 
into $W^+ W^-$ is twice the antiproton (proton) flux in our model.  

\begin{figure}[t!]\begin{center}
\includegraphics[width=6.5cm]{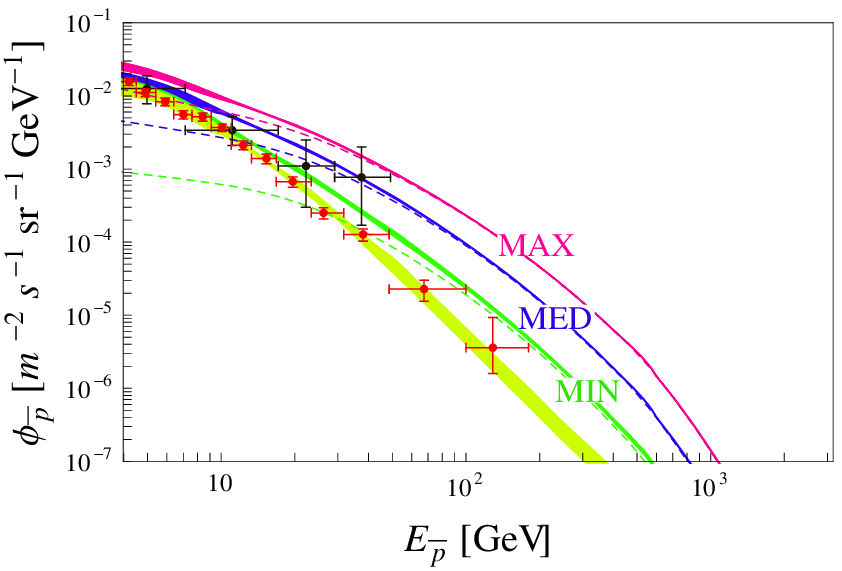}~~~~~~~~\includegraphics[width=6.5cm]{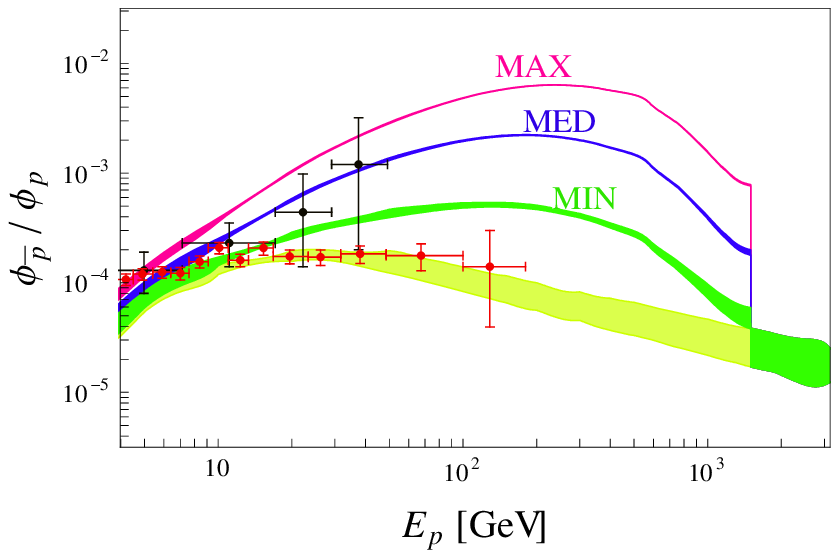}
\end{center}
\vspace*{-0.8cm} 
\caption{Antiproton flux (left) and antiproton/proton ratio (right) for the MAX, MED, MIN propagation models, 
$M_{N_1}=3$ TeV and $x_{N_1}^{1/2} C_\ell = 10^{-27.2}$. The dashed curves in the left panel show the primary 
antiprotons for MAX, MED, MIN models. The lower shaded region represents the secondary antiprotons only,
according to ref.\cite{Bringmann:2006im}. The upper three curves correspond to the sum of the primary and
secondary antiprotons. The (lower) PAMELA data \cite{Adriani:2010rc} have smaller error bars with respect to 
the (upper) CAPRICE data \cite{Boezio:2001ac}. }
\label{fig-ap}
\end{figure}

The propagation for antiprotons through the galaxy is described by a diffusion equation whose solution can be cast 
in a factorized form as discussed {\it e.g.} in \cite{Cirelli:2010xx}. In this case the astrophysical uncertainties
associated to the dark matter profile and to the propagation parameters are large, about one order of magnitude.
We display the antiproton flux in the left panel of fig. \ref{fig-ap} for the MAX, MED, MIN propagation models
and for $M_{N_1}=3$ TeV and $x_{N_1}^{1/2} C_\ell = 10^{-27.2}$. The dashed curves show these primary 
antiprotons for MAX, MED, MIN models; the lower shaded region represents the flux of the secondary 
antiprotons according to ref.\cite{Bringmann:2006im}; the upper three curves correspond to the sum of the primary and
secondary antiprotons. The (lower) PAMELA data \cite{Adriani:2010rc} have smaller error bars with respect to 
the (upper) CAPRICE data \cite{Boezio:2001ac}. 
It turns out that only the MIN propagation parameter set is compatible with the data, the MED one is barely 
compatible, while the MAX seems disfavored.

The antiproton/proton ratio is studied in the right panel of fig.\ref{fig-ap}. For the proton flux, we 
consistently adopt the parameterization of ref.\cite{Bringmann:2006im} with spectral index equal to $-2.72$.
This parameterization is valid for energies higher than about $10$ GeV. 
Again, the (lower) PAMELA data points \cite{Adriani:2010rc}
are more precise than the (upper) CAPRICE ones \cite{Boezio:2001ac}. 
The plot shows that even the MIN model displays tension
with the PAMELA data. Future measurements confirming the PAMELA results could be able to rule out the
heavy decaying neutrino model studied in this work.


\section{Conclusions}

In this work we have asked whether the electron and positron CR excesses
observed by PAMELA and Fermi-LAT can be produced by a decaying fourth lepton family Majorana neutrino $N_1$. Several more or less natural models of Majorana neutrino dark matter have been put forward in the literature in addition to the ones cited in this work. A review of different models is \cite{Fan:2010yq}.  We note, however, that our model of Majorana neutrino emerges naturally at the electroweak scale given that the new generation of leptons is needed to resolve the Witten anomaly associated to the specific technicolor extension known as Minimal Walking Technicolor \cite{Sannino:2004qp,Dietrich:2005jn,Foadi:2007ue}. We have already investigated the collider properties of our extension in \cite{Frandsen:2009fs}. There we have also discussed and compared with other models of fourth family of leptons at the electroweak scale.

This heavy neutrino mixes only with the SM charged leptons via CC interactions, so its 
primary decay products are $\ell W$, $\ell=e,\mu,\tau$,
with BR proportional to the combination $C_\ell^2/(C_e^2+C_\mu^2+C_\tau^2)$,
where $C_\ell$ represents the strength of the mixing between $N$ and the SM charged lepton $\ell$.
If the mixing angles are sufficiently small, $N_1$ is nearly stable on cosmological time scales and 
constitutes a fraction $x_{N_1}$ of the today dark matter density.

We analytically calculated the energy spectrum of the electrons and positrons produced via the decay channel 
$\ell W$ and its subsequent decays, including polarization effects. 
The shape of the energy spectrum is different for $\ell=e,\mu,\tau$,
as can be seen in fig.\ref{fig-dnde}. 

After propagation (here we adopted the approximation of \cite{Ibarra:2008qg} for
the Green function of the MED-model \cite{Delahaye:2007fr}) the contribution to the total flux from each decay
channels can be straightforwardly compared with the PAMELA and Fermi-LAT experimental data combined according to
the Sum Rules method \cite{Frandsen:2010mr}, see fig.\ref{fig-sr}. We also carried out the more standard
method of the separate comparison between the PAMELA and Fermi-LAT data in order to show the reliability of the 
SR method.   

Our results are that the model provide a plausible explanation of the observed CRs excesses,
when the BRs of $N_1\rightarrow \tau W$ and $N_1\rightarrow \mu W$ are the dominant ones, 
as it is clear from fig. \ref{fig-pf-fine}. No acceptable agreement with the data can be achieved 
for $N_1\rightarrow e W$. This means that $N_1$ should have a hierarchical pattern of mixing angles with the 
SM charged leptons, with $C_e < C_{\mu,\tau}$.
In particular, the $\tau W$ decay mode alone seems to best fit the data.
In this case, the best fit occurs for $M_{N_1} \approx 3$ TeV and $\sqrt{x_{N_1}} C_\tau \approx 10^{-27}$. 
For the $\tau W$ decay our results differ from the ones obtained in ref.\cite{Ibarra:2008qg,Meade:2009iu} 
by using numerical codes. We found that our inclusion of polarization effects cannot alone explain this
discrepancy.

We have demonstrated that the positron and electron CR excesses possibly present in the PAMELA \footnote{More recent data of the PAMELA collaboration \cite{Adriani:2011xv} on the negative electron flux has been released which, however, given the large uncertainties for energies higher than $100$~GeV do not affect our results.} and Fermi-LAT
data, can be simply accounted for in our fourth generation heavy Majorana neutrino model. 
Clearly, the contraints from gamma photons \cite{Abdo:2010ex,Abdo:2010nz} would deserve a dedicated analysis,
but we do not expect to find tension between our model and these data.
We instead estimated the primary contribution to the antiproton flux and the antiproton to proton ratio
in our model. Despite the uncertainty in the propagation parameters, it seems that even the MIN model
shows tension with the PAMELA data \cite{Adriani:2010rc}. 
The AMS-02 space station experiment will hopefully provide additional relevant informations \cite{ams02}.

\section*{Acknowledgements}

We thank M.T.Frandsen, F.Dradi, S.Hansen, T.Hapola, M.Moretti, C.A. Savoy and E.Torassa for useful discussions.


\appendix

\section{Full mixing}
\label{sec:full mixing}

The mixing between the neutrinos of flavor $\zeta$ and the three SM neutrinos is described by the $5\times 5$
mass matrix ${\cal M}$: 
\begin{equation}
-{\cal L}= \frac{1}{2} ( \begin{array}{ccccc} 
 \overline{(\nu_{e L})^c}&\overline{(\nu_{\mu L})^c}&\overline{(\nu_{\tau L})^c}& \overline{(\nu_{\zeta L})^c} & 
 \overline{\nu_{\zeta R}}  \end{array} )
\left( \begin{array}{ccc|cc}  
{\cal O}(eV) & {\cal O}(eV) &{\cal O}(eV) & {\cal O}(eV) & m_e \cr  
{\cal O}(eV) & {\cal O}(eV) &{\cal O}(eV) & {\cal O}(eV) & m_\mu \cr  
{\cal O}(eV) & {\cal O}(eV) &{\cal O}(eV) & {\cal O}(eV) & m_\tau \cr  \hline
{\cal O}(eV) & {\cal O}(eV) &{\cal O}(eV) & {\cal O}(eV) & m_D \cr 
m_e & m_\mu& m_\tau & m_D &  m_R \end{array} \right)  
\left( \begin{array}{c} \nu_{e L} \cr \nu_{\mu L} \cr \nu_{\tau L} \cr \nu_{\zeta L} \cr (\nu_{\zeta R})^c  \end{array} \right) + h.c.~~.
\label{}
\end{equation}
Barring unnatural tunings and up to corrections to its mixings of ${\cal O}(eV/M_{N_1,N_2})$,
the unitary matrix $V$ is
\begin{equation}
\left( \begin{array}{c}
\nu_{e L} \cr\nu_{\mu L} \cr\nu_{\tau L} \cr  \nu_{\zeta L} \cr  (\nu_{\zeta R})^c  \end{array} \right) = V
\left( \begin{array}{c} P_L N_e \cr  P_L N_\mu \cr P_L N_\tau \cr P_L N_1 \cr P_L N_2  \end{array} \right) ~~,~~
V=\left( \begin{array}{ccccc} 
c_e               & 0                 & 0            & i c s_e               & s s_e              \cr
-s_e s_\mu        & c_\mu             & 0            & i c c_e s_\mu         & s c_e s_\mu        \cr
-s_e c_\mu s_\tau & - s_\mu s_\tau    & c_\tau       & i c c_e c_\mu s_\tau  & s c_e c_\mu s_\tau \cr 
-s_e c_\mu c_\tau & - s_\mu c_\tau    & -s_\tau      & i c c_e c_\mu c_\tau  & s c_e c_\mu c_\tau \cr 
0                 & 0                 & 0            & -i s                  & c 
\end{array} \right)~,
\label{}
\end{equation}
where 
\bea
t_\tau^2 &=& \frac{{m}_\tau^2}{m_D^2} ~~~, ~~~~~~ t_\mu^2 = \frac{{m}_\mu^2}{ m_D^2 + {m}_\tau^2} ~~~,~~~~~
t_e^2 = \frac{{m}_e^2}{m_D^2 + {m}_\tau^2+{m}_\mu^2} ~~,\\
&~&~~~~~~\tan( 2 \theta)= 2 ~\frac{{m'}_D}{m_R} ~~,~~~{m'}_D^2=m_D^2+{m}_\tau^2+{m}_\mu^2+{m}_e^2\no,
\label{}
\eea
diagonalises the lower $2\times 2$ sector of the $5\times 5$ mass matrix ${\cal M}$, namely
\beq
V {\cal M} V^\dagger =  \left( \begin{array}{ccc}  
 m^{eff}_{3\times 3} & 0 & 0 \cr  
 0 & M_{N_1} & 0 \cr 
0  &  0 &  M_{N_2} \end{array} \right)  
\label{matnu}
\eeq
where the elements of $m^{eff}_{3\times 3}$ are naturally ${\cal O}(eV)$ and
\beq
M_{N_1}=\frac{m_R}{2} \left( \sqrt{1+4 \frac{{m'_D}^2}{m_R^2}} -1 \right) ~~,~~~~
M_{N_2}=\frac{m_R}{2} \left( \sqrt{1+4 \frac{{m'_D}^2}{m_R^2}} +1 \right) ~.
\eeq
At this stage, the matrix in eq.(\ref{matnu}) can be fully diagonalised with a further unitary matrix,
which can be identified with the MNS mixing matrix, acting only on the upper $3\times 3$ block:
\beq
U^\dagger m^{eff}_{3\times 3} U^* = {\rm diag}(m_1,m_2,m_3)~~.
\eeq
Notice that in the limit of small mixing angles between the heavy and light neutrinos, the 
mixing matrix $V$ becomes (at first order in $s_\ell$)
\beq
V\approx \left( \begin{array}{ccccc} 
1                 & 0                 & 0            & i c s_e               & s s_e              \cr
0                 & 1                 & 0            & i c s_\mu             & s s_\mu            \cr
0                 & 0                 & 1            & i c s_\tau            & s s_\tau           \cr 
-s_e              & - s_\mu           & -s_\tau      & i c                   & s                  \cr 
0                 & 0                 & 0            & -i s                  & c 
\end{array} \right)~.
\eeq

We have thus demonstrated that each light flavor couples to the heavy neutrinos as
in eq.(\ref{Vapprox}).


\end{document}